\documentclass[preprint,12pt]{elsarticle}

\usepackage{amssymb}
\usepackage{amsthm}
\usepackage{framed} 
\usepackage{amsmath,color}
\usepackage{mathrsfs}
\usepackage{graphicx}
\usepackage{epstopdf}
\usepackage{float}
\usepackage{caption}
\usepackage{subcaption}
\usepackage{bm}
\usepackage{bbm}
\usepackage{mathrsfs}
\usepackage{cleveref}
\usepackage{amssymb} 
\usepackage{extarrows} 
\usepackage{soul}
\usepackage{accents}
\usepackage{color,soul} 
\usepackage{color} 
\usepackage{tabu} 
\usepackage{longtable}
\usepackage{nomencl} 
\makenomenclature 
\biboptions{sort&compress}
\soulregister\citep7 
\soulregister\citet7 
\soulregister\citealp7 
\newsavebox{\measurebox} 
\usepackage{titlesec} 

\journal{Corrosion Science}

\makeatletter
\def\@author#1{\g@addto@macro\elsauthors{\normalsize%
    \def\baselinestretch{1}%
    \upshape\authorsep#1\unskip\textsuperscript{%
      \ifx\@fnmark\@empty\else\unskip\sep\@fnmark\let\sep=,\fi
      \ifx\@corref\@empty\else\unskip\sep\@corref\let\sep=,\fi
      }%
    \def\authorsep{\unskip,\space}%
    \global\let\@fnmark\@empty
    \global\let\@corref\@empty  
    \global\let\sep\@empty}%
    \@eadauthor={#1}
}
\makeatother

\titleformat{\paragraph}
{\normalfont\normalsize\itshape}{\theparagraph}{1em}{}
\titlespacing*{\paragraph}
{0pt}{3.25ex plus 1ex minus .2ex}{1.5ex plus .2ex}

\begin{document}

\begin{frontmatter}



\title{Generalised boundary conditions for hydrogen transport at crack tips}


\author{Emilio Mart\'{\i}nez-Pa\~neda\corref{cor1}\fnref{Cam}}
\ead{e.martinez-paneda@imperial.ac.uk}

\author{Andres D\'{\i}az\fnref{UBU}}

\author{Louise Wright\fnref{NPL}}

\author{Alan Turnbull\fnref{NPL}}

\address[Cam]{Department of Civil and Environmental Engineering, Imperial College London, London SW7 2AZ, UK}

\address[UBU]{Structural Integrity Group, University of Burgos, Escuela Polit\'{e}cnica Superior. Avenida Cantabria s/n, 09006, Burgos, Spain}

\address[NPL]{National Physical Laboratory, Teddington, Middlesex, TW11 0LW, UK}

\cortext[cor1]{Corresponding author. Tel: +45 45 25 42 71; fax: +45 45 25 19 61.}

\begin{abstract}
We present a generalised framework for resolving the electrochemistry-diffusion interface and modelling hydrogen transport near a crack tip. The adsorption and absorption kinetics are captured by means of Neumann-type generalised boundary conditions. The diffusion model includes the role of trapping, with a constant or evolving trap density, and the influence of the hydrostatic stress. Both conventional plasticity and strain gradient plasticity are used to model the mechanical behaviour of the solid. Notable differences are found in the estimated crack tip hydrogen concentrations when comparing with the common procedure of prescribing a constant hydrogen concentration at the crack surfaces.
\end{abstract}


\end{frontmatter}



\begin{framed}
\nomenclature{$\theta_L, \, \theta_r$}{fractional occupancy in lattice and trapping sites}
\nomenclature{$N_L, \, N_r$}{density of lattice and trapping sites}
\nomenclature{$C, \, C_r$}{hydrogen concentration in lattice and trapping sites}
\nomenclature{$t$}{time}
\nomenclature{$R$}{universal gas constant}
\nomenclature{$T$}{temperature}
\nomenclature{$\bar{V}_H$}{partial molar volume of hydrogen}
\nomenclature{$\sigma_h$}{hydrostatic stress}
\nomenclature{$k_r$}{rate constant for capture to trapping sites}
\nomenclature{$p_r$}{rate constant for release to lattice sites}
\nomenclature{$M$}{metal atom interacting with hydrogen at surface}
\nomenclature{$H_{ads}$}{atomic solute hydrogen in adsorption sites}
\nomenclature{$H_{abs}$}{atomic solute hydrogen in absorption sites}
\nomenclature{$k_{abs}$}{rate constant for absorption}
\nomenclature{$k_{des}$}{rate constant for desorption}
\nomenclature{$C_s$}{sub-surface hydrogen concentration}
\nomenclature{$\theta_{ad}, \, \theta_{ad}^R$}{surface coverage and reference surface coverage at 1 atm}
\nomenclature{$N_s$}{density of absorption sites}
\nomenclature{$J_{in}$}{absorption flux}
\nomenclature{$\mathcal{S}$}{surface boundary}
\nomenclature{$\mu_{L}, \, \mu_{H_2}$}{interstitial lattice and $H_2$ chemical potentials}
\nomenclature{$\mu_{L}^0, \, \mu_{H_2}^0$}{reference interstitial lattice chemical and $H_2$ chemical potentials}
\nomenclature{$f_{H_2}$}{fugacity}
\nomenclature{$p^0$}{pressure in the standard state.}
\nomenclature{$K, \, K_0$}{solubility constant and pre-exponential of the solubility constant}
\nomenclature{$E_s$}{activation energy for gaseous dissociation}
\nomenclature{$\eta$}{overpotential}
\nomenclature{$F$}{Faraday constant}
\nomenclature{$A$}{constant related to the HER kinetic parameters}
\nomenclature{$\zeta$}{constant related to the HER kinetic parameters}
\nomenclature{HER}{Hydrogen Evolution Reaction}
\nomenclature{$k_v, \, k_t, \, k_h$}{forward reaction rates for Volmer, Tafel and Heyrovsky reactions}
\nomenclature{$\alpha_v, \, \alpha_h$}{charge transfer coefficients for Volmer and Heyrovsky reactions}
\nomenclature{$\Gamma$}{surface concentration of adsorption sites}
\nomenclature{$i_c$}{charging current density}
\nomenclature{$i_{r,chem}, \, i_{r,elec}$}{chemical and electrochemical recombination current densities}
\nomenclature{$k_c$}{charging current constant}
\nomenclature{$k_{r,chem}, \, k_{r,elec}$}{chemical  and electrochemical recombination constants}
\nomenclature{$\sigma_y$}{initial yield stress}
\nomenclature{$\nu$}{Poisson's ratio}
\nomenclature{$K_I$}{mode I stress intensity factor}
\nomenclature{$r_p$}{plastic zone size}
\nomenclature{$r_0$}{crack tip radius}
\nomenclature{$r_b$}{outer radius of boundary layer model}
\nomenclature{$E_B$}{trap binding energy}
\nomenclature{$E$}{Young's modulus}
\nomenclature{$D_L$}{lattice diffusion coefficient}
\nomenclature{$\varepsilon_p$}{equivalent plastic strain}
\nomenclature{$b_0, \, b$}{initial and current crack tip opening displacement}
\nomenclature{$N$}{strain hardening exponent}
\nomenclature{$u, \, v$}{horizontal and vertical displacement components}
\nomenclature{$\Psi$}{free energy}
\nomenclature{$\varepsilon_{ij}^e$}{elastic strain tensor}
\nomenclature{$\varepsilon_{ij}^p$}{plastic strain tensor}
\nomenclature{$C_{ijkl}$}{isotropic elastic stiffness tensor}
\nomenclature{$\mu$}{shear modulus}
\nomenclature{$E^p$}{generalised effective plastic strain}
\nomenclature{$L_E, \, L_D, \, \ell$}{energetic, dissipative and reference material length scales}
\nomenclature{$k_r^0, \, p_r^0$}{pre-exponential constants for capture and release rates}
\nomenclature{$E^t, \, E^d$}{activation energies for capture and release}
\nomenclature{$E_L$}{activation energy for lattice diffusion}

\printnomenclature
\end{framed}

\section{Introduction}
\label{Sec:Introduction}

Hydrogen assisted cracking is an important problem for a wide range of metals in aqueous environments. While the underlying mechanisms are not completely understood \cite{Nagao2018,Lynch2019,JMPS2020,Shishvan2020}, experiments consistently show a notable reduction in fracture toughness and fatigue resistance with increasing hydrogen content \cite{Gangloff2003,Gangloff2012}. Independently of the mechanisms at play, damage is related to the hydrogen concentration in the fracture region and its quantification is important in determining the likelihood of cracking. Accordingly, the analysis of hydrogen transport near a crack tip has received significant attention \cite{Sofronis1989,Krom1999,Dadfarnia2011,Barrera2016,Diaz2016b,IJHE2016}. The study of hydrogen transport comprises both the bulk transport, i.e. lattice diffusion and trapping phenomena, and the surface-related processes that govern hydrogen entry from the environment, i.e. the adsorption-absorption mechanisms.\\

Hydrogen transport deviates from conventional Fickian diffusion through various mechanisms, but near a crack tip mainly through trapping and stress-driven diffusion. A numerical framework considering both effects was established by Sofronis and McMeeking  \cite{Sofronis1989}. Their pioneering work is based on finite strain J2 plasticity theory and incorporates the following diffusion features: (i) effect of trapping, trapped hydrogen delays diffusion from ideal lattice behaviour \textit{via} a sink term in the mass balance; (ii) Oriani's equilibrium, the derivation of this trapping term assumes equilibrium as proposed by Oriani \cite{Oriani1972}; (iii) stress-driven diffusion, the chemical potential decreases with increasing hydrostatic stress, following thermodynamic arguments \cite{Li1966}; and (iv) trap density dependence on plastic straining, as inferred from the permeation tests by Kumnick and Johnson \cite{Kumnick1980}. These features translate into a peak lattice hydrogen concentration at a certain distance ahead of the crack, coinciding with the hydrostatic stress peak, and a concentration of trapped hydrogen at the crack tip. The modelling assumptions by Sofronis and McMeeking \cite{Sofronis1989} hold in certain regimes but key questions and phenomena remain to be addressed. Particularly important are: (a) the modelling of hydrogen entry, (b) the validity of equilibrium, and (c) the role of crack tip dislocation hardening mechanisms. The present work evaluates these three aspects, with special focus on the development of generalised boundary conditions that are able to mimic hydrogen entry from a wide range of environmental conditions. Other effects, such as self-stresses \cite{Johnson2003}, are not considered but could be relevant for high solubility metals such as Ni-based alloys \cite{Li2017} or those that form hydrides \cite{Lufrano1998a}. Also, we assume spherical dilatation and leave unaddressed the influence of tetragonal distortion \cite{McLellan1970,Zhang1992,Sanchez2008}. Considering the steps of the Hydrogen Evolution Reaction (HER), the variation of hydrogen coverage is related to the adsorption/desorption phenomena and to the charging input variables, i.e. to the charging current and to the overpotential. The concept of fugacity is also discussed with the aim of evaluating the differences between constant concentration and a generalised flux as boundary conditions. Critical transition parameters and important scaling relationships related to hydrogen transport in the crack environment and reactions at the crack surfaces are identified and quantified. \\

Using a Prandtl stress field, Turnbull \emph{et al.} \cite{Turnbull1996} compared the cases of prescribing a constant hydrogen concentration at the crack faces and the use of a generalised flux as boundary condition, showing notable differences in the hydrogen distribution near the crack tip. Despite these results, the vast majority of hydrogen transport studies published to date still rely on the use of Dirichlet-type constant concentration boundary conditions, most likely due to the simplicity of its numerical implementation. We here provide a robust numerical framework for generalised flux boundary conditions that can be coupled to not only a Prandtl solid but to any constitutive model, under both small and large strain conditions. In addition, we also consider a kinetic approach for the relationship between trapped hydrogen and lattice concentration and revisit the work by Turnbull \emph{et al.} \cite{Turnbull1996} under the conditions simulated by Sofronis and McMeeking \cite{Sofronis1989}. The aim is to illustrate the differences between both modelling strategies in realistic scenarios, delimit the regimes of applicability, and draw conclusions for hydrogen-related failures. In addition, the hypothesis that hydrogen distribution is independent of the initial crack tip opening is assessed within a general discussion on a critical distance for hydrogen assisted cracking.\\ 

The present work also aims at gaining insight into the coupling of this generalised framework with the influence of plastic flow near crack tips, shedding light into the competition between surface-electrochemical effects and plasticity-enhanced solubility. The extra storage of dislocations required to accommodate lattice curvature due to non-uniform plastic deformation leads to high crack tip stresses that conventional plasticity models are unable to capture \cite{Ashby1970,Komaragiri2008,IJP2016}. This crack tip stress elevation associated with Geometrically Necessary Dislocations (GNDs) and dislocation hardening mechanisms brings a notable increase in the lattice hydrogen concentration in the fracture process zone \cite{AM2016}. Modelling this phenomenon, in combination with generalised boundary conditions, is expected to give quantitative insight into the spatial and time scales involved during hydrogen assisted cracking. 


\section{Hydrogen transport model}
\label{Sec:Theory}

Hydrogen atoms can occupy normal interstitial lattice sites and can also reside at trapping sites, such as interfaces or dislocations. The hydrogen concentration in the lattice can be defined as,
\begin{equation}
    C = \theta_L N_L
\end{equation}

\noindent where $N_L$ denotes the number of interstitial sites per unit volume and $\theta_L$ is the lattice occupancy fraction ($0<\theta_L<1$). All trapping sites are considered reversible, which is effectively the case if a sufficiently wide range of time scales and temperatures is considered. Thus, the hydrogen concentration at reversible traps is denoted by $C_r$, and is given by,
\begin{equation}
    C_r = \theta_r N_r
\end{equation}

\noindent where $N_r$ is the reversible trap density and $\theta_r$ is the fractional occupancy of reversible trap sites.\\

The local mass conservation of lattice hydrogen concentration $C$ and reversibly trapped hydrogen concentration $C_r$ is given by an extended version of Fick's second law, as:
\begin{equation}\label{Eq:Htransport}
    \frac{\partial C}{\partial t} + \frac{\partial C_r}{\partial t} = D_L \nabla^2 C + \nabla \left(- \frac{D_L C}{R T} \bar{V}_H \nabla \sigma_h \right)
\end{equation}

\noindent where $D_L$ is the lattice diffusion coefficient, $\bar{V}_H$ is the partial molar volume of hydrogen atoms, $\sigma_h$ is the hydrostatic stress, $R$ is the gas constant, and $T$ is the absolute temperature. Following Turnbull and co-workers \cite{Turnbull1996,Turnbull1997}, capture and release are explicitly simulated by taking into account the kinetic formulation first proposed by McNabb and Foster \cite{McNabb1963}. Thus, the variation of trapped hydrogen is defined as
\begin{equation}\label{Eq:mcnabb_CT}
    \frac{\partial C_r}{\partial t} = N_r \left[ k_r C \left( 1 - \theta_r \right) - p_r \theta_r \right]
\end{equation}

\noindent Here, $k_r$ and $p_r$ are rate constants for capture to trapping sites and release to lattice sites, respectively. Unless otherwise stated, we generally follow Turnbull \textit{et al.} \cite{Turnbull1997} in assuming a constant trap density. Thus, the rate of trapping can be expressed in terms of occupancy,
\begin{equation}
    \frac{\partial C_r}{\partial t} = N_r \frac{\partial \theta_r}{\partial t}
\end{equation}

We make use of the finite element method to discretise and solve the hydrogen transport equation (\ref{Eq:Htransport}), with the lattice hydrogen concentration $C$ being the primary kinematic variable. This is coupled with the solution of the following differential equation for $\theta_r$,
\begin{equation}\label{Eq:thetaR}
    \frac{\partial \theta_r}{\partial t} = k_r C (1 - \theta_r ) - p_r \theta_r
\end{equation}

It must be noted that, following Ref. \cite{Turnbull1993}, $k_r$ and $p_r$ take different units. A different nomenclature can be adopted by which a constant, $k_r^* = k_rN_L$, can be defined with the same units as $p_r$, expressing Eq. (\ref{Eq:thetaR}) in terms of lattice occupancy rather than concentration:
\begin{equation}
    \frac{\partial \theta_r}{\partial t} =k_r^* \theta_L \left( 1 - \theta_r \right) - p \theta_r
\end{equation}


\section{Generalised Boundary Conditions}
\label{Sec:BC}

Setting the ground for modelling the stages of hydrogen entry, the theory on surface effects for hydrogen-metal interaction is reviewed in this section. Mechanical analyses, required to characterise crack tip fields, are rarely enriched with models from electrochemistry science; this is despite the Hydrogen Evolution Reaction (HER) being one of the most widely studied electrochemical processes \cite{Lasia2010}.\\

Local damage and hydrogen accumulation within the Fracture Process Zone (FPZ) are influenced by hydrogen entry from an aqueous solution, which depends on the reaction mechanisms operating at metallic surfaces. 
From a numerical consideration, the governing diffusion equation must be supplied with appropriate boundary conditions, which are related to the adsorption and absorption phenomena. Two modelling strategies are usually adopted: prescribing a constant concentration (Dirichlet boundary conditions) or prescribing a constant normal flux (Neumann boundary conditions). It is often assumed that potentiostatic charging produces a constant surface concentration while galvanostatic charging can be modelled by a constant entry flux \cite{Conway1994,Pumphrey1980}. However, these two ideal scenarios are unlikely to be attained due to the role of the finite rate constants present in the absorption-desorption process, $k_{abs}$ and $k_{des}$: 
\begin{equation}\label{Eq:ads_abs_reaction}
    MH_{ads}  \xleftrightarrow[k_{des}]{k_{abs}} MH_{abs}
\end{equation}

The absorption flux can be formulated in terms of the rate constants $k_{abs}$ and $k_{des}$. Taking into consideration that a flux of hydrogen atoms is required to reach equilibrium between adsorbed sites and sub-surface concentration, Pumphrey \cite{Pumphrey1980} defined the absorption flux $J_{in}$ as,
\begin{equation}\label{Eq:JabsPumprey}
       J_{in}=k_{abs} \theta_{ad} - k_{des} C_s
\end{equation}

\noindent where $\theta_{ad}$ is the surface coverage fraction and $C_s$ is the sub-surface concentration. Here, the absorption constant $k_{abs}$ has the same units as the flux, mol/(m$^{2}\cdot$s) or ppm$\cdot$m/s, whereas the desorption constant $k_{des}$ has SI units of m/s. This equation has been subsequently adopted in many studies; see, for example, Refs. \cite{Zhang1999a,Turnbull1993}. Eq. (\ref{Eq:JabsPumprey}) constitutes an appropriate simplification of the absorption reaction for the case of $\theta_{ad} \ll 1$ and low surface concentration, i.e. $C_s \ll N_s$, where $N_s$ is the number of absorption sites per unit volume. A more general definition is given as follows:
\begin{equation}\label{Eq:JabsExtended}
       J_{in}=k_{abs}^* (N_s-C_s) \theta_{ad} - k_{des} C_s(1-\theta_{ad} )
\end{equation}

\noindent where the absorption rate constant has been redefined as $k_{abs}^*$ and is given now in the same units as $k_{des}$. Once the flux has been defined, an expression for the surface coverage $\theta_{ad}$ can be obtained from the adsorption behaviour of the hydrogen-metal interface.\\

In the present modelling framework, see Section \ref{Sec:Theory}, the concentration of absorbed hydrogen in surface sites, $C_s$, corresponds to the lattice concentration $C$ at the boundary $\mathcal{S}$. Thus, the absorption sites take lattice variables and the following equivalence is assumed: 
\begin{equation}\label{Eq:C_s}
       \frac{C_s}{N_s} = \frac{C(\mathcal{S})}{N_L} = \theta_L(\mathcal{S})
\end{equation}

\noindent The absorption flux can then be reformulated as:
\begin{equation}\label{Eq:Jabs_thetaL}
       J_{in}=k_{abs}^* N_L(1-\theta_L) \theta_{ad} - k_{des} C (1-\theta_{ad} )
\end{equation}

\noindent where lattice quantities $C$ and $\theta_L$ are determined at the boundary. A limiting case might be defined when absorption and desorption constants are large in comparison to the input flux so $J_{in}/k_{abs}$ tends to zero; in that case, a relationship between bulk occupancy and surface coverage can be established,
\begin{equation}\label{Eq:surface_equil}
      \frac{\theta_L}{1-\theta_L} = \frac{k_{abs}^*}{k_{des}}\frac{\theta_{ad}}{1-\theta_{ad}}
\end{equation}

We restrict our attention to iron-based alloys, in which the solubility is low and the concentration in lattice sites is significantly smaller than the number of interstitial locations, i.e. $\theta_L \ll 1$. Accordingly, the lattice concentration at the surface reads:
\begin{equation}\label{Eq:surface_concentr}
      C (S) = \frac{k_{abs}^*N_L}{k_{des}}\frac{\theta_{ad}}{1-\theta_{ad}}=\frac{k_{abs}}{k_{des}}\frac{\theta_{ad}}{1-\theta_{ad}}
\end{equation}

\noindent The low occupancy assumption also simplifies the absorption flux expression:
\begin{equation}\label{Eq:Jabs}
       J_{in}=k_{abs}\theta_{ad} - k_{des} C (1-\theta_{ad} )
\end{equation}

An alternative approach for determining an equilibrium subsurface concentration, without involving the coverage $\theta_{ad}$, is based on the concept of fugacity and the equivalence to gaseous charging. Under equilibrium conditions, the chemical potential of $H_2$, i.e. $\mu_{H_2}$, and that of the interstitial hydrogen, $\mu_{L}$, are related as:
\begin{equation}\label{Eq:chem_pot_eq}
       \mu_{L} =\frac{1}{2} \mu_{H_2} 
\end{equation}

\noindent Each term can be expanded considering the corresponding chemical activities:
\begin{equation}\label{Eq:chem_pot_eq1}
       \mu_{L}^0+ RT\ln{\frac{C}{N_L}} =\frac{1}{2}\mu_{H_2}^0 + RT \ln{\sqrt{\frac{f_{H_2}}{p^0}}}
\end{equation}

\noindent where $\mu_{H_2}^0$ is the reference $H_2$ chemical potential and low occupancy, $\theta_L \ll 1$, is assumed. The fugacity $f_{H_2}$ is defined in relation to the pressure in the standard state $p^0$, which is usually taken as $10^5$ Pa. Even though the number of lattice sites remains constant, their chemical potential is reduced by the hydrostatic stress \cite{Li1966}:
\begin{equation}\label{Eq:chem_pot}
       \mu_{L}^{\sigma} =\mu_{L}-\bar{V}_H \sigma_h 
\end{equation}

Including the hydrostatic stress term in Eq. (\ref{Eq:chem_pot_eq1}) and rearranging, an equilibrium concentration can be obtained as:
\begin{equation}\label{Eq:chem_pot_eq2}
       C = \frac{N_L}{\sqrt{p^0}}\exp{\left (-\frac{\mu_L^0-\frac{1}{2}\mu_{H_2}^0}{RT}\right)}
       \exp{\left (\frac{\bar{V}_H}{RT}\sigma_h\right)}\sqrt{f_{H_2}}
\end{equation}

Eq. (\ref{Eq:chem_pot_eq2}) is a generalisation of the typical Sievert's law; the stress influence is accounted for and pressure is substituted by fugacity. The Arrhenius nature of solubility $K$ is demonstrated, with $K_0 = N_L/\sqrt{p^0}$ being the pre-exponential term while the activation energy for gaseous dissociation is given by the term $E_s = \mu_L^0-\mu_{H_2}^0/2$. Di Leo and Anand \cite{DiLeo2013} showed that adopting the chemical potential as primary kinematic variable in the mass transport problem can naturally capture the stress-dependent boundary condition. This scheme has also been recently adopted by Elmukashfi \textit{et al.} \cite{Elmukashfi2020}. A constant surface chemical potential can also be prescribed in the context of a model where lattice concentration is the primary kinematic variable, as done by D\'{\i}az \textit{et al.} \cite{Diaz2016b} and Mart\'{\i}nez-Pa\~neda \textit{et al.} \cite{IJHE2016}.\\

Sievert's law is commonly used to obtain the boundary concentration under gaseous charging conditions, i.e. the concentration is proportional to the square root of hydrogen partial pressure \cite{Sofronis1989}. In order to establish an equivalence with absorption from a $H_2$ gaseous environment, absorption and adsorption constants can be reformulated as fugacity. Thus, defining $\theta_{ad}^R$ as the surface hydrogen coverage at 1 atm and considering the absorption-adsorption process:
\begin{equation}\label{Eq:surface_equil2}
      \frac{\theta_{ad}}{1-\theta_{ad}} = \frac{\theta_{ad}^R}{1-\theta_{ad}^R}\sqrt{f_{H_2}}
\end{equation}

\noindent Expressions for the fugacity can be obtained at steady state conditions by considering the complete Hydrogen Evolution Reaction (HER), i.e. the Volmer-Heyrovsky-Tafel reactions \cite{Bockris1971,Liu2014}. Generally, the fugacity is related to the overpotential $\eta$ (a negative quantity) via an Arrhenius function \cite{Liu2014}:
\begin{equation}\label{Eq:fugacity_QianLiu}
      f_{H_2}=A\exp{\left(-\frac{\eta F}{\zeta RT}\right)}
\end{equation}

\noindent where $A$ and $\zeta$ are constants, which are related to the kinetic parameters involved in the HER. An alternative is to treat them as empirical constants, to be fitted to permeation tests. In acid solutions, the Hydrogen Evolution Reaction is given by the following three steps \cite{Harrington1987}: 
\begin{align*}
&\text{Adsorption:}          &   H^+ + M + e^-  &\xleftrightarrow[-k_{v}]{k_{v}} MH_{ads} \\
&\text{Chemical recombination:}         &  2MH_{ads}  &\xleftrightarrow[-k_{t}]{k_{t}} H_2 + 2M\\
&\text{Electrochemical recombination:}   &  MH_{ads} + M^+ + e^-  &\xleftrightarrow[-k_{h}]{k_{h}} H_2 + M
\end{align*}

Following Liu \textit{et al.} \cite{Liu2014}, and neglecting the terms corresponding to backward reactions, the coverage evolution can be explicitly modelled as:
\begin{equation}\label{Eq:coverage_qianliu}
\begin{split}
\frac{\partial \theta_{ad}}{\partial t} & =  2 k_v C_{H^+} (1-\theta_{ad}) \exp \left(-\alpha_v\frac{\eta F}{RT}\right) \\
 & - k_t \theta_{ad}^2\\
& -2 k_h C_{H^+} \theta_{ad} \exp \left(-\alpha_h\frac{\eta F}{RT}\right) 
\end{split}
\end{equation}

\noindent where $k_v$, $k_t$ and $k_h$ are the forward reaction rate constants for Volmer, Tafel and Heyrovsky reactions, respectively. The charge transfer coefficients ($\alpha_v$ and $\alpha_h$) are only involved in the electrochemical Volmer and Heyrovsky steps. The input flux is related with the coverage rate \cite{Montella1999}, and can be divided in three currents:
\begin{equation}\label{Eq:Jin_currents}
      J_{in}=\Gamma \frac{\partial \theta_{ad}}{\partial t} =\frac{1}{F} (i_c + i_{r,chem} + i_{r,elec})
\end{equation}

\noindent where $\Gamma$ is the Surface concentration of adsorption sites (mol/m$^{2}$). And, following Turnbull \textit{et al.} \cite{Turnbull1993}, reaction constants can be grouped as: 
\begin{equation}\label{Eq:ic}
i_c =  2 \Gamma F k_v C_{H^+} (1-\theta_{ad}) \exp \left(-\alpha_v\frac{\eta F}{RT}\right) = Fk_c (1-\theta_{ad})
\end{equation}
\begin{equation}\label{Eq:ir_chem}
i_{r,chem} =  - \Gamma F k_t \theta_{ad}^2 = -Fk_{r,chem} \theta_{ad}^2
\end{equation}
\begin{equation}\label{Eq:ir_elec}
i_{r,elec} =  - 2 \Gamma F k_h C_{H^+} \theta_{ad} \exp \left(-\alpha_h\frac{\eta F}{RT}\right) = -Fk_{r,elec} \theta_{ad}
\end{equation}

Thus, $k_c$ and $k_{r,elec}$ depend on electrical overpotential $\eta$ and on the concentration $C_{H^+}$, i.e. on the pH. Assuming constant overpotential, pH and temperature, the adsorption flux can be simplified to:
\begin{equation}\label{Eq:Jads}
    J_{in}=k_c \left( 1 - \theta_{ad} \right) - k_{r,chem} \theta^2_{ad}- k_{r,elec} \theta_{ad}
\end{equation}

Eq. (\ref{Eq:Jads}) is the generalised boundary condition that is prescribed at the crack surfaces in the present numerical framework. Considering that the adsorption flux is much smaller than the charging current constant, i.e. $J_{in}/k_c$ tends to zero, the coverage value is constant and might be found by imposing equation (\ref{Eq:Jads}) equal to zero. The reaction constants can take different quantities at the crack wall and the crack tip. By equating (\ref{Eq:Jabs}) and (\ref{Eq:Jads}) one reaches a relationship between the sub-surface concentration $C$ and the coverage $\theta_{ad}$. The latter is readily obtained for every time point, without the need of assuming a small flux, by solving the second-order equation:
\begin{equation}\label{Eq:theta_ad}
    k_{r,chem} \theta_{ad}^2 + \left( k_{abs} \exp \left(\frac{\bar{V}_H\sigma_h}{RT} \right) + k_{des} C + k_c + k_{r,elec} \right) \theta_{ad} - k_{des} C - k_c= 0
\end{equation}


\section{Results}
\label{Sec:Results}

The formulation described in Sections \ref{Sec:Theory} and \ref{Sec:BC} is implemented into a finite element framework, and subsequently employed to showcase model predictions and gain physical insight. First, the numerical implementation is described and validated against results from the literature in Section \ref{Sec:VerificationTurnbull1996} and Appendices A and B. Secondly, in Section \ref{Sec:AISI4340steel}, the model is used to quantify the influence of generalised boundary conditions and rate constants on hydrogen behaviour in AISI 4340 steel, following Ref. \cite{Turnbull2015}. Then, we mimic the paradigmatic benchmark of Sofronis and McMeeking \cite{Sofronis1989} in a model iron-based material (Section \ref{Sec:SofronisMcMeeking}). The influence of McNabb-Foster kinetics and generalised boundary conditions is investigated. These two material systems are then used to investigate the role of crack tip opening (Section \ref{Sec:Influencecrack}), trap density (Section \ref{Sec:InfluenceNr}) and local crack tip strain gradient strengthening (Section \ref{Sec:StrainGradientPlasticity}).

\subsection{Numerical implementation and verification}
\label{Sec:VerificationTurnbull1996}

The finite element framework is developed and validated in a rigorous step-by-step strategy. The hydrogen transport model with McNabb-Foster kinetics is addressed first, in the absence of generalised boundary conditions and mechanical deformation. Thus, the theoretical framework described in Section \ref{Sec:Theory} is implemented by solving Eqs. (\ref{Eq:Htransport}) and (\ref{Eq:thetaR}) in a coupled manner. We choose as primary kinematic variables, and nodal degrees of freedom, the lattice hydrogen concentration $C$ and the trap occupancy $\theta_r$. As detailed in \ref{App:TDS}, model predictions are benchmarked against the Thermal Desorption Spectroscopy (TDS) analysis of Legrand \textit{et al.} \cite{Legrand2015}, showing a perfect agreement. The second step involves the implementation, in the absence of mechanical loading, of the generalised boundary conditions in the McNabb-Foster hydrogen transport model. Thus, a Neumann-type boundary condition is prescribed based on Eq. (\ref{Eq:Jabs}), with $\theta_{ad}$ being an internal variable that depends on the solution, as given by Eq. (\ref{Eq:theta_ad}). The framework is validated against the simulations by Turnbull and co-workers \cite{Turnbull2014,Turnbull2015} of stress-free permeation where electrochemical surface conditions govern hydrogen uptake, see \ref{App:EP}. Finally, the complete framework is developed, in what constitutes the first finite element implementation of a coupled mechanical-diffusion model based on McNabb-Foster kinetics and the first numerical model solving the mechanical problem coupled to generalised boundary conditions. The system is composed of the mechanical force balance, the mass transport balance (\ref{Eq:Htransport}), and the trapping kinetics equation (\ref{Eq:thetaR}). Displacements, lattice hydrogen concentration and trap occupancy are the primary variables. In addition to the standard boundary conditions, a flux-type boundary condition is prescribed based on Eq. (\ref{Eq:Jabs}). Details of the validation are described below.\\

We validate the complete framework by addressing the crack problem considered by Turnbull \textit{et al.} \cite{Turnbull1996}. Specifically, we aim at quantitatively reproducing the effect of the trapping rate constant $k_r$ on the crack tip hydrogen distribution. As in Ref. \cite{Turnbull1996}, we assume that the mechanical behaviour of the solid is given by a Prandtl stress field. Thus, for a polar coordinate system ($r,\theta$) centred at the crack tip and assuming plane strain conditions, the hydrostatic stress in the plastic region ($r \leq r_p$) is given by,
\[ \sigma_h =
  \begin{cases}
    \sigma_y / \sqrt{3}       & \quad \text{if } 3 \pi /4< | \theta | < \pi \\
     \sigma_y / \sqrt{3}  \left( 1 + \frac{3 \pi}{2} - 2 \theta \right)  & \quad \text{if } \pi /4< | \theta | < 3 \pi/4\\
    \sigma_y / \sqrt{3}  \left( 1 +  \pi \right)  & \quad \text{if } | \theta | < \pi/4
  \end{cases}
\]

\noindent where $\sigma_y$ is the material yield stress. Outside of the plastic zone ($r>r_p$), the stress field is given as a function of the applied mode I stress intensity factor $K_I$ by the linear elastic solution:
\begin{equation}
    \sigma_h = \frac{2 (1 + \nu)}{3 \sqrt{2 \pi r}} K_I \cos \frac{\theta}{2}
\end{equation}

\noindent with $\nu$ being Poisson's ratio. The size of the plastic zone, $r_p$, is defined as the location where the elastic field and the Prandtl field coincide.\\

\begin{figure}[H]
  \makebox[\textwidth][c]{\includegraphics[width=1\textwidth]{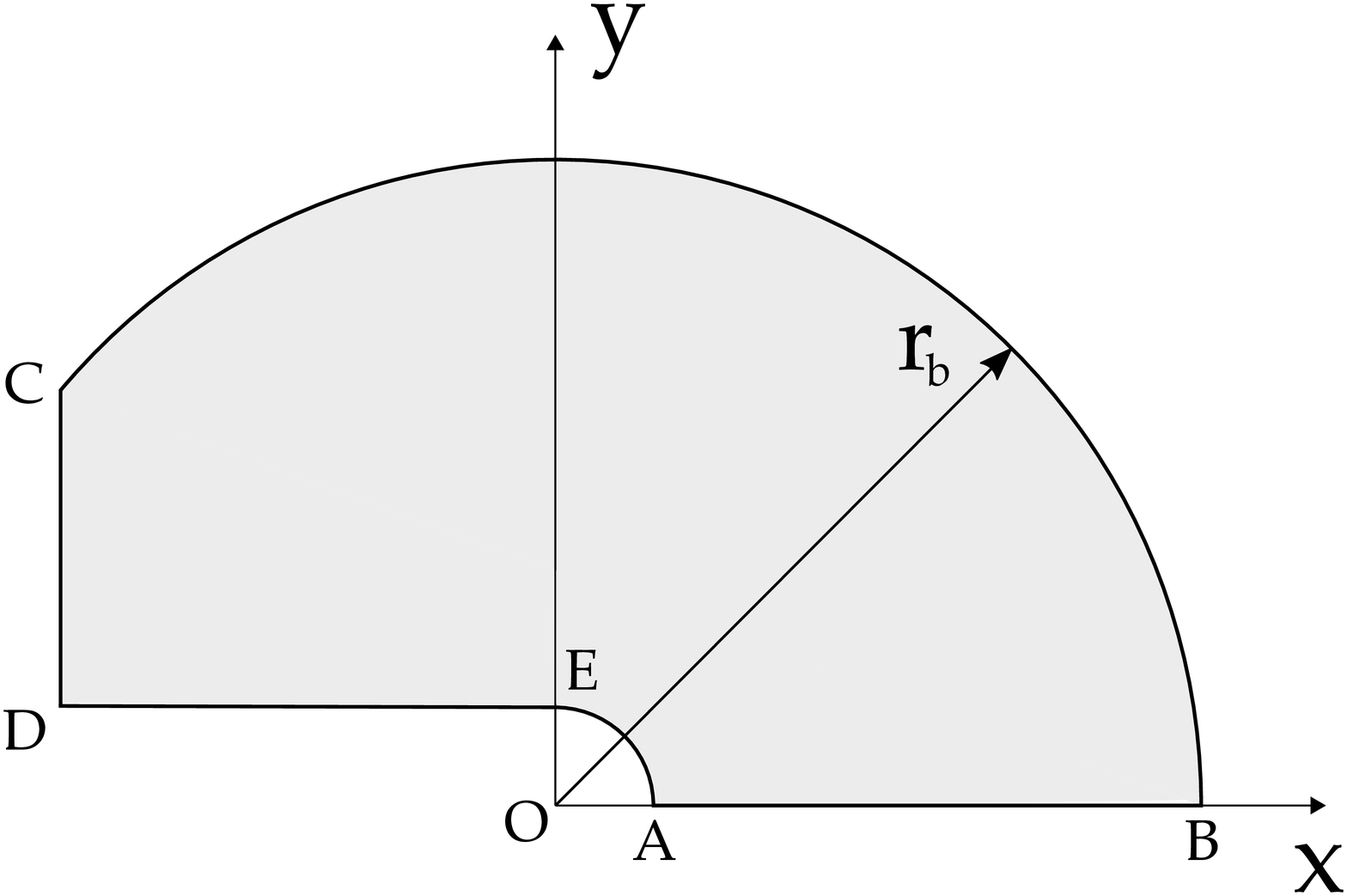}}%
  \caption{Influence of generalised boundary conditions with a Prandtl stress field, validation with the results from Ref. \cite{Turnbull1996}. Sketch of the boundary value problem.}
  \label{fig:MeshGeometryCrackVerification}
\end{figure}

The geometry and configuration of the problem are shown in Fig. \ref{fig:MeshGeometryCrackVerification}. We follow Turnbull \textit{et al.} \cite{Turnbull1996} and consider a remote load of $K_I=30$ MPa$\sqrt{m}$ and a crack tip radius of $r_0 =0.98$ $\mu$m (EO). A very refined mesh is employed near the crack tip, with the characteristic element size being equal to 0.2 $\mu$m. As in Ref. \cite{Turnbull1996}, it is assumed that the hydrogen concentration at $t=0$ is equal to $C=0$ in the entire specimen. In addition, we prescribe the Neumann-type boundary condition expressed in (\ref{Eq:Jads}) but neglecting electrochemical recombination, i.e. $k_{r,elec}=0$, on the crack wall (DE),
\begin{equation}
    J_{in}^w = k_c^w \left( 1 - \theta_{ad}^w \right) - k_{r,chem}^w (\theta_{ad}^w)^2
\end{equation}

\noindent and on the crack tip (EA),
\begin{equation}
    J_{in}^t = k_c^t \left( 1 - \theta_{ad}^t \right) - k_{r,chem}^t (\theta_{ad}^t)^2
\end{equation}

\noindent where $\theta_{ad}^w$ and $\theta_{ad}^t$ are the surface coverages of hydrogen atoms on the crack wall and crack tip, respectively, as computed from (\ref{Eq:theta_ad}). In addition, $k_c^w \left( 1 - \theta_{ad}^w \right)$ and $k_c^t \left( 1 - \theta_{ad}^t \right)$ represent the current densities for reduction of hydrogen ions at the crack walls and tip, respectively, divided by Faraday's constant. And $k_{r,chem}^w$ and $k_{r,chem}^t$ are the hydrogen atom recombination rate constants for the crack wall and tip. The diffusion, mechanical and geometrical parameters employed are given in Table \ref{Tab:case0}. The ratio $k_r/p_r$ is fixed while the capture constant $k_r$ is varied to explore the sensitivity of crack tip hydrogen distributions. This relationship between capture and release constants, due to their respective dependences on trapping and detrapping energies, depends on the binding energy of traps, $E_B$:
\begin{equation}
    \frac{k_r}{p_r} = \frac{k_r^*/N_L}{p_r} = \frac{1}{N_L}\exp{\left(\frac{E_B}{RT}\right)}
\end{equation}

\begin{table}[H]
\centering
\caption{Diffusion, mechanical and geometrical parameters for the verification study, following Ref. \cite{Turnbull1996}.}
\label{Tab:case0}
   {\tabulinesep=1.2mm
   \makebox[\textwidth][c]{\begin{tabu} {cccccccccccccc}
       \hline
$D_L$  & $N_r$ & $k_r/p_r$ & $E_B$ & $N_L$ & $C_0$ \\ \hline
7.2$\times$10$^{-9}$ & 2.2$\times$10$^{24}$  & 1.1$\times$10$^{-21}$ & 49.0 & 4.95$\times$10$^{29}$ & 0 \\
(m$^2$/s) & (sites/m$^3$) &  (m$^3$/site) & (kJ/mol) & (sites/m$^3$) & (wt ppm)  
   \end{tabu}}}
 \\
      {\tabulinesep=1.2mm
   \makebox[\textwidth][c]{\begin{tabu} {cccccccccccc}
       \hline
$T$ & $\bar{V}_H$ & $\sigma_y$  & $\nu$ & $r_0$ \\ \hline
293 & 2$\times$10$^{-6}$ & 1200 & 0.3 & 0.98\\
(K) & (m$^3$/mol) & (MPa) & (-) & ($\mu$m)  \\\hline
   \end{tabu}}}
\end{table}

The constants related to the absorption/desorption and adsorption processes that are employed for the crack tip and wall are given in Table \ref{Tab:case0walltip}, following Ref. \cite{Turnbull1996}. Since the charging constant $k_c^t$ is considered 10 times higher than its wall counterpart $k_c^w$, it is expected that hydrogen entry will be enhanced near the crack tip and that the influence of the hydrostatic stress will be magnified.\\

\begin{table}[H]
\centering
\caption{Parameters related to the boundary conditions for the crack wall and the crack tip in the verification study, following Ref. \cite{Turnbull1996}.}
\label{Tab:case0walltip}
   {\tabulinesep=1.2mm
   \makebox[\textwidth][c]{\begin{tabu} {ccccc}
       \hline
& $k_{abs}$ [mol/(s $\cdot$m$^2$)]  & $k_c$ [mol/(s $\cdot$m$^2$)] & $k_{des}$ [m/s] & $k_{r,chem}$ [mol/(s $\cdot$m$^2$)]\\ \hline
Crack wall, $k_i^w$ & 1$\times$10$^{11}$ & 5$\times$10$^{-7}$ & 8.8$\times$10$^{9}$ & 22\\
Crack tip, $k_i^t$ & 1$\times$10$^{11}$ & 5$\times$10$^{-6}$ & 8.9$\times$10$^{9}$ & 22\\\hline
   \end{tabu}}}
\end{table}

\begin{figure}[H]
  \makebox[\textwidth][c]{\includegraphics[width=1\textwidth]{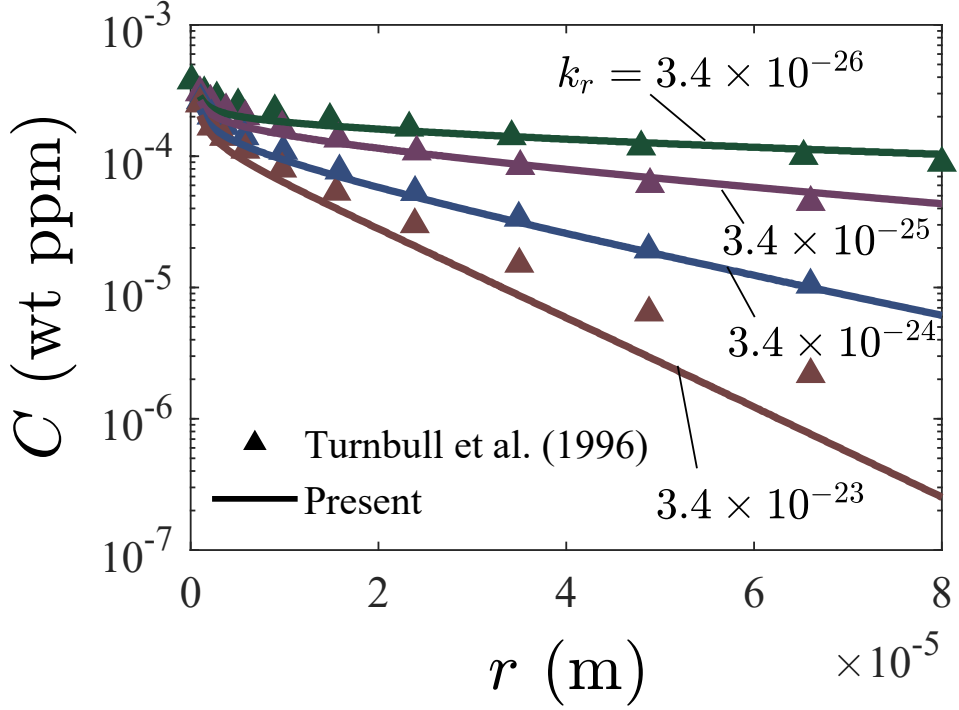}}%
  \caption{Validation with the results from \cite{Turnbull1996}. Effect of the trapping rate constant $k_r$ on the crack tip lattice hydrogen distribution for $K_I=30$ MPa$\sqrt{m}$ at a time of $t=67$ s. The units of $k_r$ are m$^3$s$^{-1}$site$^{-1}$.}
  \label{fig:Turnbull0}
\end{figure}

The computed lattice hydrogen distributions ahead of the crack tip are shown in Fig. \ref{fig:Turnbull0} for several $k_r$ values. The results agree reasonably well with those by Turnbull \textit{et al.} \cite{Turnbull1996} despite the different numerical methodology and coarser mesh employed in their study. In agreement with expectations, the hydrogen concentration increases with decreasing $k_r$ due to the slower trap filling rate.

\subsection{First case study: AISI 4340 Steel}
\label{Sec:AISI4340steel}

\begin{figure}[H]
  \makebox[\textwidth][c]{\includegraphics[width=1\textwidth]{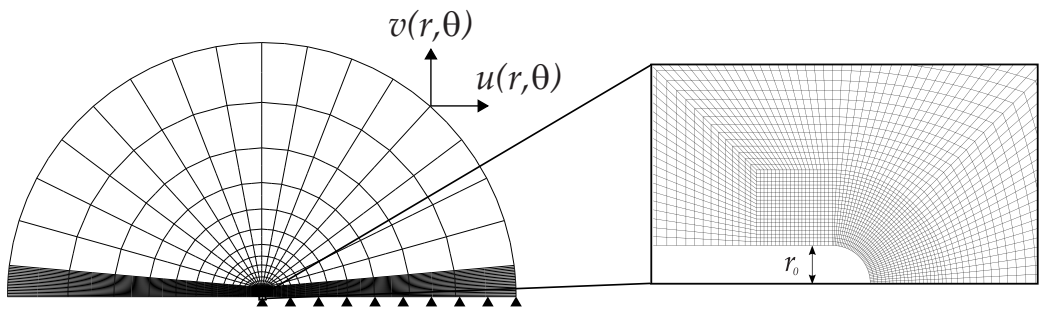}}%
  \caption{General and detailed representation of the finite element mesh employed for the boundary layer model. Mechanical boundary conditions are shown superimposed.}
  \label{fig:MeshCase1}
\end{figure}

Once validated, the modelling framework is extended to characterise the mechanical response by means of finite strain J2 plasticity. The first case study aims at assessing the role of generalised boundary conditions on AISI 4340 steel under a more realistic choice of material model. We follow the work by Sofronis and McMeeking \cite{Sofronis1989} and make use of the so-called boundary layer formulation, with the crack tip being blunted with a radius $r_0=5$ $\mu m$, see Fig. \ref{fig:MeshCase1}. Taking advantage of symmetry, only half of the specimen is modelled and a remote $K_I$ field is imposed by prescribing the displacements at the outer radius of the mesh, $r_b$. The mesh is refined in the region near the crack tip, with the characteristic element size being equal to $r_0/12$. Small scale yielding conditions are assumed and the ratio $r_b/r_0$ equals 30000. For a polar coordinate system centered at the crack tip, the outer periphery of the mesh ($r=r_b$) is subjected to the mode I elastic $K_I$-field by prescribing the following horizontal $u$ and vertical $v$ nodal displacements,
\begin{equation}
u \left( r, \theta \right) = K_I \frac{1+\nu}{E} \sqrt{\frac{r}{2 \pi}} \cos \left( \frac{\theta}{2} \right) \left(3 - 4 \nu - \cos \theta \right)
\end{equation}
\begin{equation}
v \left( r, \theta \right) = K_I \frac{1+\nu}{E} \sqrt{\frac{r}{2 \pi}} \sin \left( \frac{\theta}{2} \right) \left(3 - 4 \nu - \cos \theta \right)
\end{equation}

\noindent where $E$ is Young's modulus. Work hardening is captured by means of the following isotropic power law,
\begin{equation}
    \sigma = \sigma_y \left( 1 + \frac{E \varepsilon_p}{\sigma_y} \right)^N
\end{equation}

\noindent where $\varepsilon_p$ is the effective plastic strain and $N$ is the strain hardening exponent. In this case study, we adopt the mechanical properties for AISI 4340 Steel given by Turnbull \textit{et al.} \cite{Turnbull1996}, and assume that $E=207$ GPa and $N=0.2$, see Table \ref{Tab:mechcase1}.

\begin{table}[H]
\centering
\caption{Mechanical parameters for the first case study, AISI 4340 Steel, following Ref. \cite{Turnbull1996}.}
\label{Tab:mechcase1}
   {\tabulinesep=1.2mm
   \makebox[\textwidth][c]{\begin{tabu} {ccccc}
       \hline
 $K$ & $\sigma_y$ & $E$ & $\nu$ & $N$  \\ \hline
 30 & 1200 & 207000 & 0.3 & 0.2 \\
 (MPa$\sqrt{m}$) & (MPa) & (MPa) & (-) & (-) \\\hline
   \end{tabu}}}
\end{table}

The diffusion and absorption/adsorption parameters follow Ref. \cite{Turnbull1996} and the verification case study of Section \ref{Sec:VerificationTurnbull1996}; i.e., the parameters are given in Tables \ref{Tab:case0} and \ref{Tab:case0walltip}. Since the influence of pre-charging was shown to be relatively small in Ref. \cite{Turnbull1996}, we assume no pre-charging $C_0=0$ mol/m$^3$. The constant concentration model assumes small fluxes in the absorption reaction, i.e. $J_{in}/k_{abs} = 0$ in (\ref{Eq:Jabs_thetaL}), and relates surface concentration to coverage \textit{via} Eq. (\ref{Eq:surface_concentr}). Since the adsorption flux is also assumed to achieve very small values after a long time, a constant coverage $\theta_{ad}$ can be calculated by imposing $J_{in}/k_c = 0$ in (\ref{Eq:Jads}). This latter assumption gives $\theta_{ad}^t = 4.77 \times$10$^{-4}$ and $\theta_{ad}^w = 1.51 \times$10$^{-4}$. The corresponding constant concentrations are $C^t =5.42 \times$10$^{-3}$ mol/m$^3$ and $C^w = 1.72 \times$10$^{-3}$ mol/m$^3$. The former value, $C^t$, corresponds to the $6.9 \times 10^{-4}$ wt ppm magnitude considered in \cite{Turnbull1996}.\\

\begin{figure}[H]
  \makebox[\textwidth][c]{\includegraphics[width=1\textwidth]{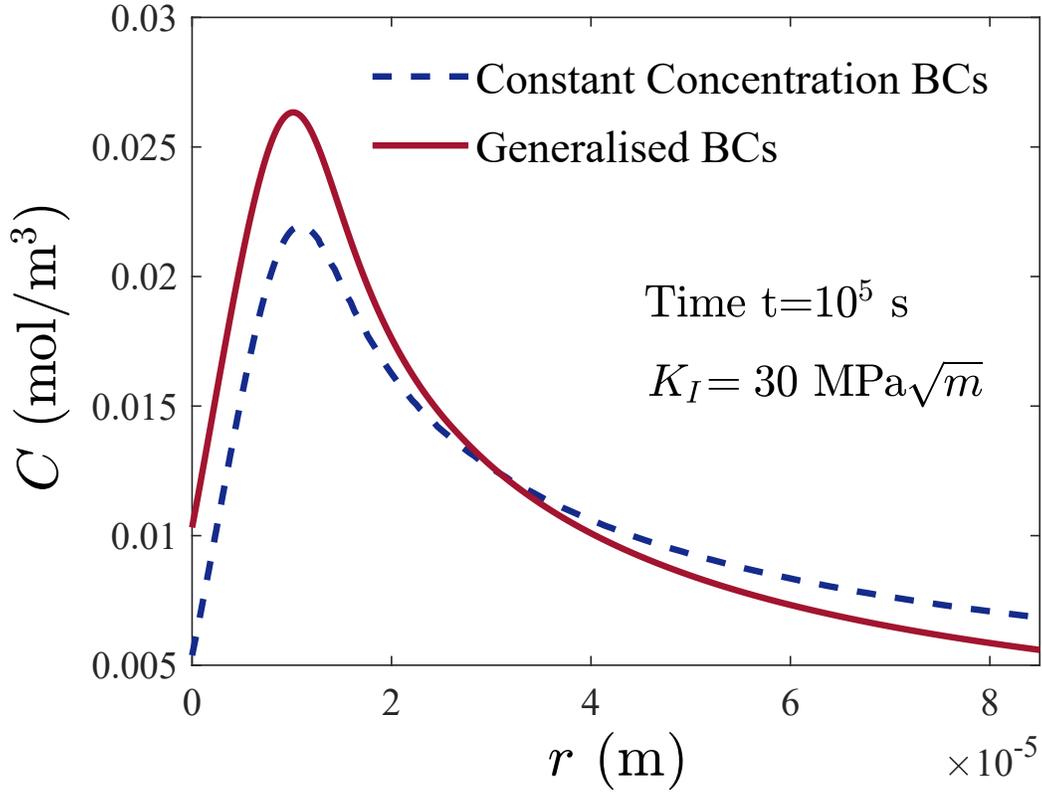}}%
  \caption{First case study, AISI 4340 steel. Generalised boundary conditions versus constant concentration boundary conditions. Hydrogen distribution ahead of the crack for a load of $K_I=30$ MPa$\sqrt{m}$ and a total time of $10^5$ s. Trapping rate constant $k_r=3.4 \times 10^{-26}$ m$^3$s$^{-1}$site$^{-1}$.}
  \label{fig:Fig1K30GFvsCC}
\end{figure}

We address first the differences between different boundary conditions. Results are shown in Fig. \ref{fig:Fig1K30GFvsCC} for a remote load of $K_I=30$ MPa$\sqrt{m}$. A large time scale is considered, such that the solution is expected to be close to that of steady state. The generalised boundary conditions lead to a higher hydrogen concentration at the crack tip and a larger peak, relative to the commonly used constant hydrogen concentration scheme. As expected, the Constant Concentration (CC) model shows a surface concentration of $5.42 \times$10$^{-3}$ mol/m$^3$. The larger hydrogen concentration attained at the crack tip is due to the $\sigma_h$-dependence of the flux boundary conditions, and differences will therefore increase with the remote load $K_I$. We emphasize that constant concentration models can be modified to account for the influence of $\sigma_h$ on the crack tip hydrogen concentration \cite{IJHE2016,Diaz2016b}. The influence of other constitutive models, such as strain gradient plasticity, will be evaluated later on.

\begin{figure}[H]
        \begin{subfigure}[h]{1.1\textwidth}
                \centering
                \includegraphics[scale=0.85]{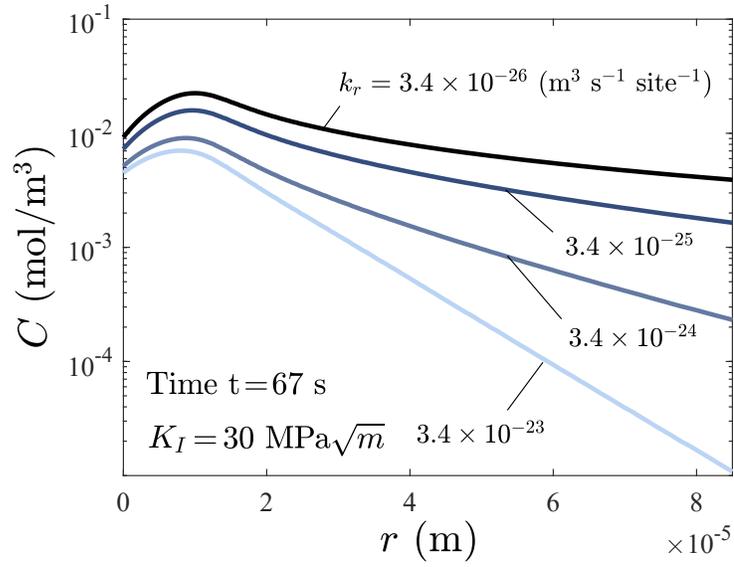}
                \caption{}
                \label{fig:Fig2K30GF_kr}
        \end{subfigure}\\
		
        \begin{subfigure}[h]{1.1\textwidth}
                \centering
                \includegraphics[scale=0.85]{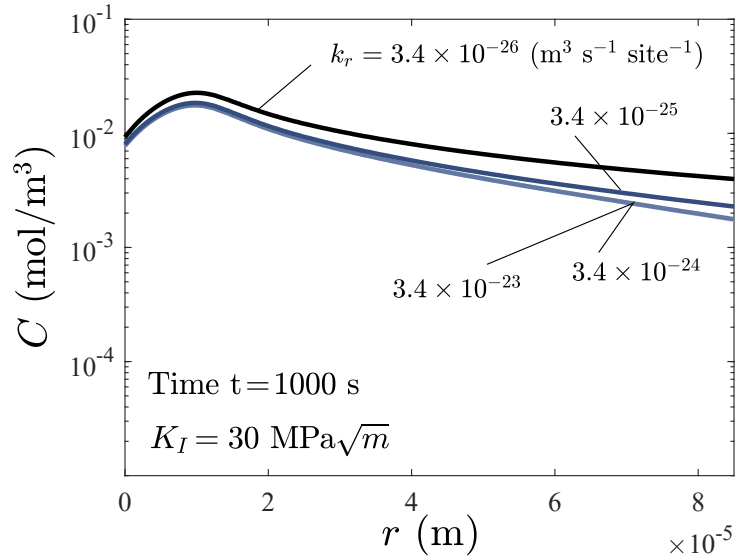}
                \caption{}
                \label{fig:Fig2K30GF_krC}
        \end{subfigure}       
        \caption{First case study, AISI 4340 steel. Influence of the trapping rate constant. Hydrogen distribution ahead of the crack for a load of $K_I=30$ MPa$\sqrt{m}$ for different values of $k_r$: (a) time $t=67$ s, and (b) time $t=1000$ s.}\label{fig:Fig2K30GF}
\end{figure}

We also investigate the influence of the trapping rate constant on the hydrogen distribution ahead of the crack tip, and its dependence with time, see Fig. \ref{fig:Fig2K30GF}. First, for a time of $t=67$ s, the sensitivity of the hydrogen distribution to $k_r$ is shown in Fig. \ref{fig:Fig2K30GF_kr}. In agreement with expectations, the hydrogen concentration increases with decreasing $k_r$ because a slower trapping process is being simulated. The sensitivity to $k_r$ decreases with time, as shown in Fig. \ref{fig:Fig2K30GF_krC} for $t=1000$ s.

\subsection{Second case study: Iron-based model material}
\label{Sec:SofronisMcMeeking}

We proceed now to investigate the influence of generalised boundary conditions in an iron-based material by reproducing the paradigmatic benchmark by Sofronis and McMeeking \cite{Sofronis1989}. The same boundary value problem as in Section \ref{Sec:AISI4340steel} is considered. A crack opening displacement $b$ is defined, such that $b_0=2r_0$, and we follow the same normalisation as Ref. \cite{Sofronis1989}: the distance to the crack tip is normalised by the crack opening displacement ($r/b$) and the hydrostatic stress is normalised by the yield stress ($\sigma_h / \sigma_y$). In the Oriani-based, constant hydrogen concentration analysis of Ref. \cite{Sofronis1989} the generalised boundary parameters intrinsic to the present framework are absent, so the values adopted for the AISI 4340 steel study are considered. The diffusion and kinetic parameters adopted are listed in Table \ref{Tab:diffcase2}.

\begin{table}[H]
\centering
\caption{Electrochemical parameters for the second case study, model iron-based material, following Ref. \cite{Sofronis1989} and Section \ref{Sec:AISI4340steel}.}
\label{Tab:diffcase2}
   {\tabulinesep=1.2mm
   \makebox[\textwidth][c]{\begin{tabu} {cccccccccccc}
       \hline
$D_L$  & $T$ & $\bar{V}_H$ & $k_r/p_r$ & $E_B$ & $N_L$ & $C_0$ \\ \hline
$1.27 \times$10$^{-8}$ & 300 & $2.0 \times$10$^{-6}$ & $5.48 \times$10$^{-20}$ & 60.0 & $5.1 \times$10$^{29}$ & $3.46 \times$10$^{-3}$ \\
(m$^2$/s) & (K) & (m$^3$/mol) & (m$^3$/site) & (kJ/mol) & (sites/m$^3$) & (mol/m$^3$)  \\\hline
   \end{tabu}}}
\end{table}

In contrast with the first case study, the number of trapping sites is defined as a function of the equivalent plastic strain, $\varepsilon_p$, to mimic the analysis by Sofronis and McMeeking \cite{Sofronis1989}. The relation follows the experimental results by Kumnick and Johnson \cite{Kumnick1980}:
\begin{equation}
    \log{N_r} = 23.26-2.33 \exp({-5.5 \varepsilon_p})
\end{equation}

\noindent Moreover, the trap binding energy ($E_B = 60$ kJ/mol) and the temperature ($T=300$ K) differ from the first case study, so the ratio $k_r/p_r$ also changes. The number of interstitial sites per unit volume is estimated assuming tetrahedral site occupancy, as appropriate for bcc iron. Following Ref. \cite{Sofronis1989}, the mechanical response is governed by finite strain conventional plasticity and the material parameters assumed for an iron-based material are given in Table \ref{Tab:mechcase2}. A remote load of $K_I=89$ MPa$\sqrt{m}$ is applied. 

\begin{table}[H]
\centering
\caption{parameters for the second case study, model iron-based material, following Ref. \cite{Sofronis1989}.}
\label{Tab:mechcase2}
   {\tabulinesep=1.2mm
   \makebox[\textwidth][c]{\begin{tabu} {cccc}
       \hline
 $\sigma_y$ & $E$ & $\nu$ & $N$  \\ \hline
 250 & 207000 & 0.3 & 0.2 \\
 (MPa) & (MPa) & (-) & (-) \\\hline
   \end{tabu}}}
\end{table}

\begin{figure}[H]
  \makebox[\textwidth][c]{\includegraphics[width=1.1\textwidth]{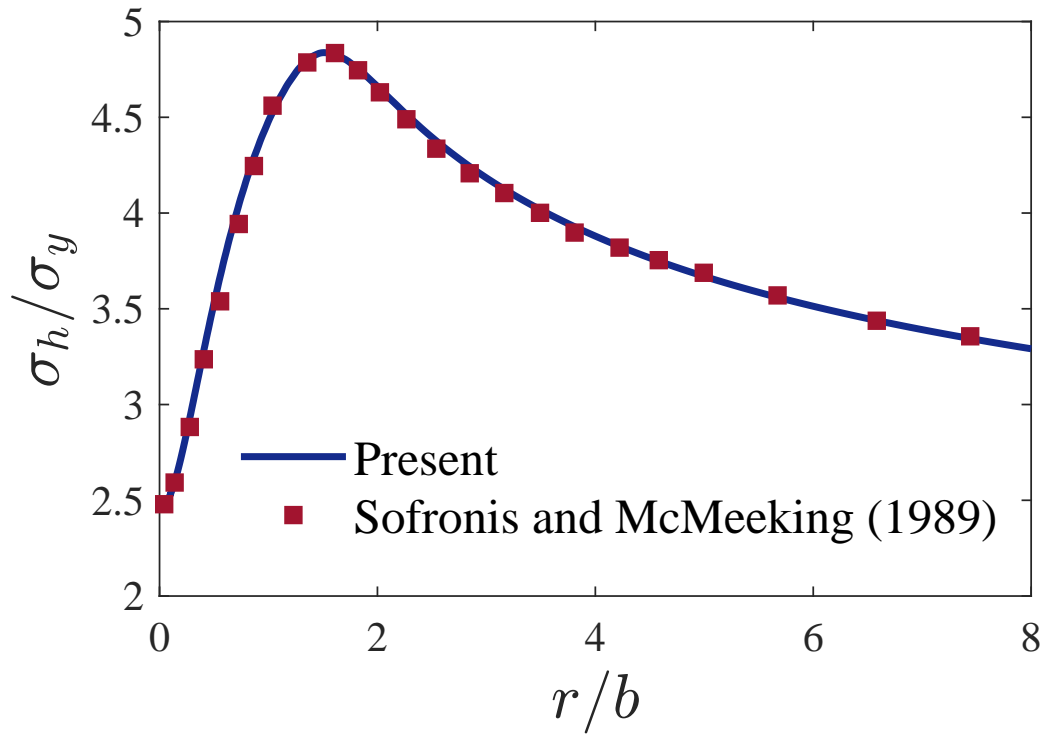}}%
  \caption{Second case study, model iron-based material. Hydrostatic stress distribution ahead of the crack tip for a load of $K_I=89$ MPa$\sqrt{m}$. Comparison with the results from Sofronis and McMeeking \cite{Sofronis1989}.}
  \label{fig:Fig1ShvsR}
\end{figure}

First, the crack tip hydrostatic stress distribution for $K_I=89$ MPa$\sqrt{m}$ is shown in Fig. \ref{fig:Fig1ShvsR}, along with the result of Sofronis and \cite{Sofronis1989}. In agreement with expectations, the same mechanical behaviour is predicted.

\begin{figure}[H]
  \makebox[\textwidth][c]{\includegraphics[width=1.2\textwidth]{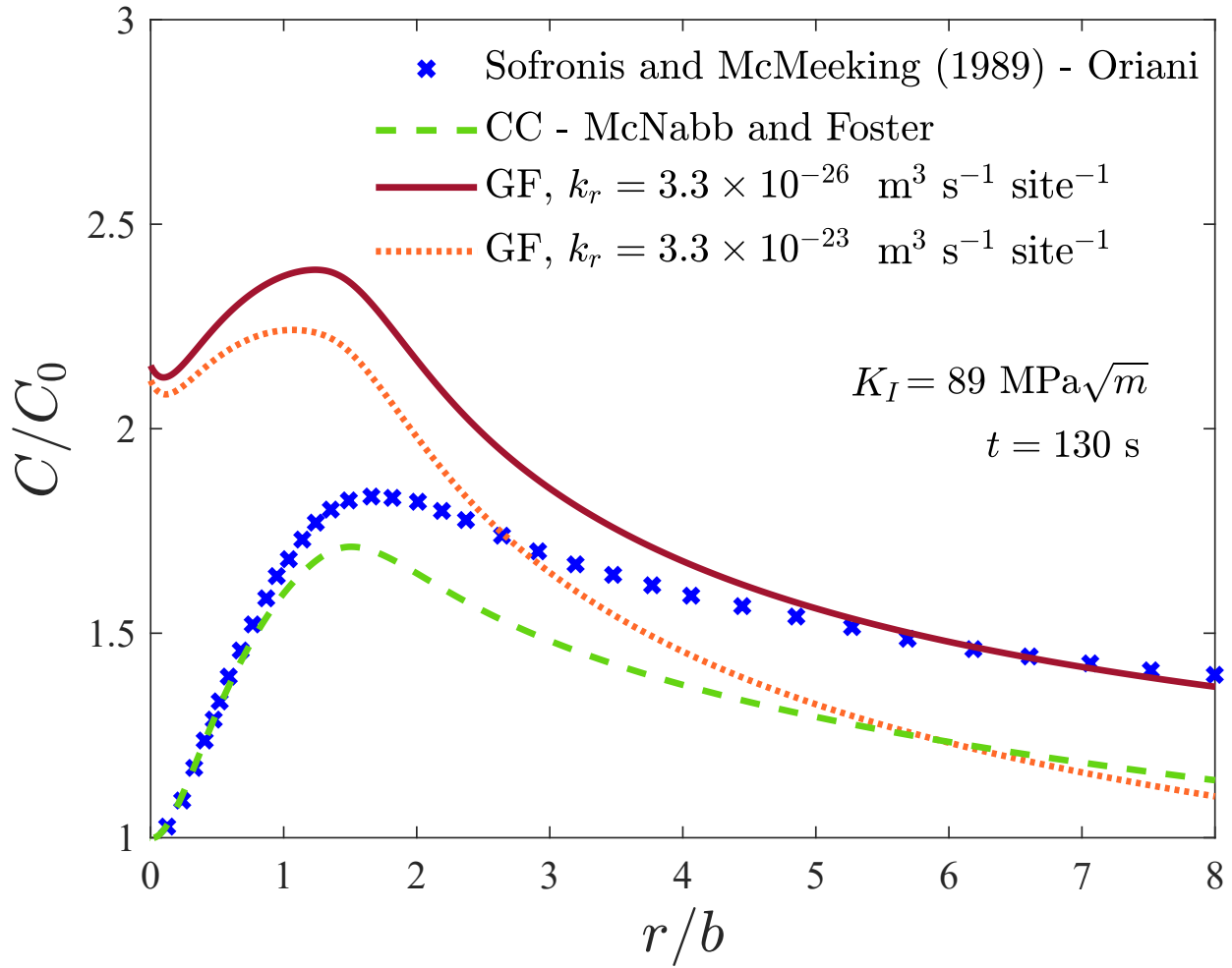}}%
  \caption{Second case study, model iron-based material. Hydrogen distribution ahead of the crack tip at 130 s.}
  \label{fig:Fig2CCvsGF}
\end{figure}

We then proceed to compute the crack tip hydrogen distribution at a time of $t=130$ s. Results are shown in Fig. \ref{fig:Fig2CCvsGF}, with the hydrogen concentration normalised by the initial hydrogen concentration $C_0$, which in Ref. \cite{Sofronis1989} coincides with the hydrogen concentration prescribed at the crack tip. The magnitude is taken to be equal to $2.084 \times$10$^{21}$ hydrogen atoms per m$^3$ (i.e. $3.46 \times$10$^{-3}$ mol/m$^3$ or $4.34 \times$10$^{-4}$ wt ppm), as in the original reference. All the lattice hydrogen concentration distributions are normalised by this magnitude. Results are obtained for three cases: (i) a constant hydrogen concentration at the crack tip equal to $C_0$, and generalised boundary conditions with (ii) $k_r=3.3 \times 10^{-26}$ m$^3$s$^{-1}$site$^{-1}$ and (iii) $k_r=3.3 \times 10^{-23}$ m$^3$s$^{-1}$site$^{-1}$. 

Consider first the results obtained with a constant hydrogen concentration (CC). Noticeable differences are shown relative to the results by Sofronis and McMeeking \cite{Sofronis1989} as the distance to the crack tip increases. These differences are due to the use of McNabb-Foster, as opposed to Oriani; when Oriani's equilibrium is enforced in our framework, the results are identical to those obtained by Sofronis and McMeeking \cite{Sofronis1989}. A smaller concentration peak is predicted when the kinetics of hydrogen trapping are resolved. To the best of the authors' knowledge, the influence of McNabb-Foster kinetics on this paradigmatic benchmark not been addressed before.\\

\begin{figure}[H]
  \makebox[\textwidth][c]{\includegraphics[width=1\textwidth]{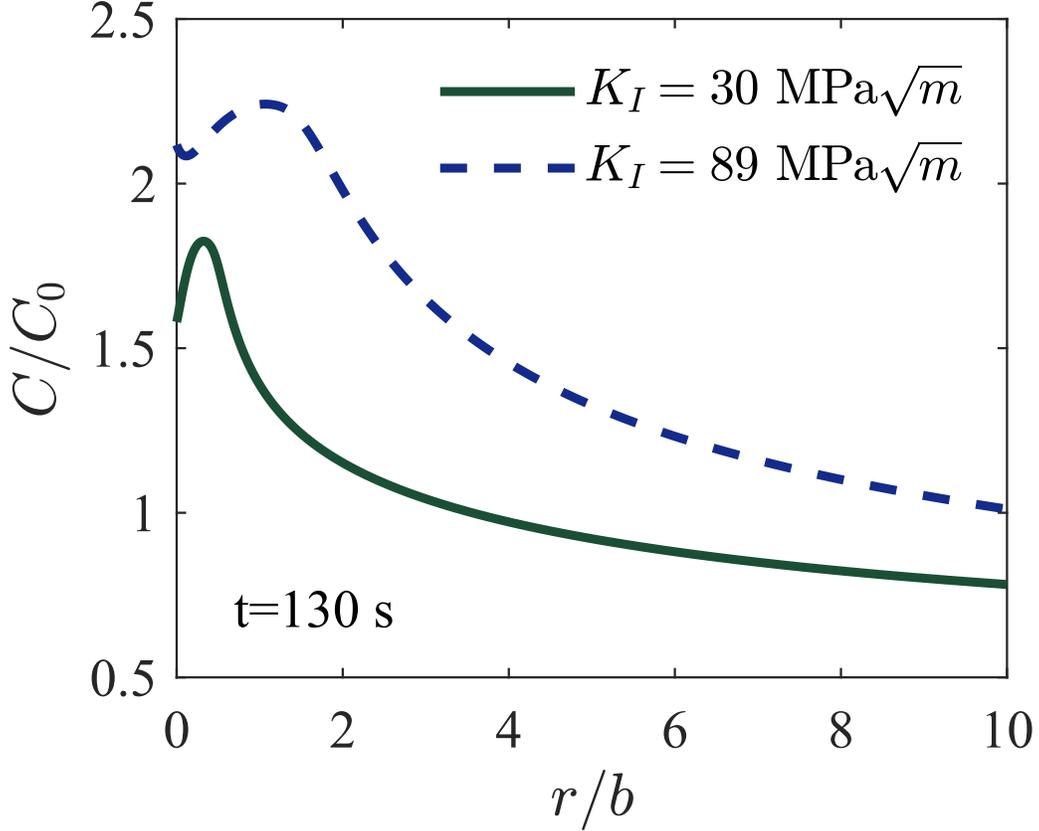}}%
  \caption{Second case study, model iron-based material. Influence of the remote load on crack tip hydrogen concentration using generalised boundary conditions; trapping rate constant $k_r=3.3 \times 10^{-23}$ m$^3$ s$^{-1}$ site$^{-1}$.}
  \label{fig:Fig3_K}
\end{figure}

Consider now the results obtained when adopting generalised flux (GF) boundary conditions with the assumed absorption/adsorption constants. Significantly larger hydrogen concentrations are predicted close the crack tip for the two values of $k_r$ considered. The result is due to the effect of the hydrostatic stress and is also inherently related to the choices of the constants $k_{abs}$, $k_{des}$, $k_{c}$ and $k_{r,chem}$. A parametric study on the influence of these constants is performed in \ref{App:k_influence}, where the evolution of sub-surface concentration and input flux is plotted versus time. Figures shown in \ref{App:k_influence} also show that $J_{in}$ approximates zero and $C_s$ remains constant after a certain time which depends on the absorption/adsorption parameters. Predictions from GF and CC modelling approaches are only expected to be equivalent after this \emph{surface-dominated} initial period. The magnitude of this surface-dominated period could have particularly important implications in environmentally assisted fatigue \cite{EFM2017}. Finally, the influence of the remote load $K_I$ is investigated in Fig. \ref{fig:Fig3_K} using generalised boundary conditions. In agreement with expectations, the crack tip hydrogen concentration shows sensitivity to the value of $K_I$ and the hydrogen distribution increases with the applied load.

\subsection{Influence of the crack tip opening}
\label{Sec:Influencecrack}

\begin{figure}[H]
\makebox[\linewidth][c]{%
        \begin{subfigure}[b]{0.53\textwidth}
                \centering
                \includegraphics[scale=0.6]{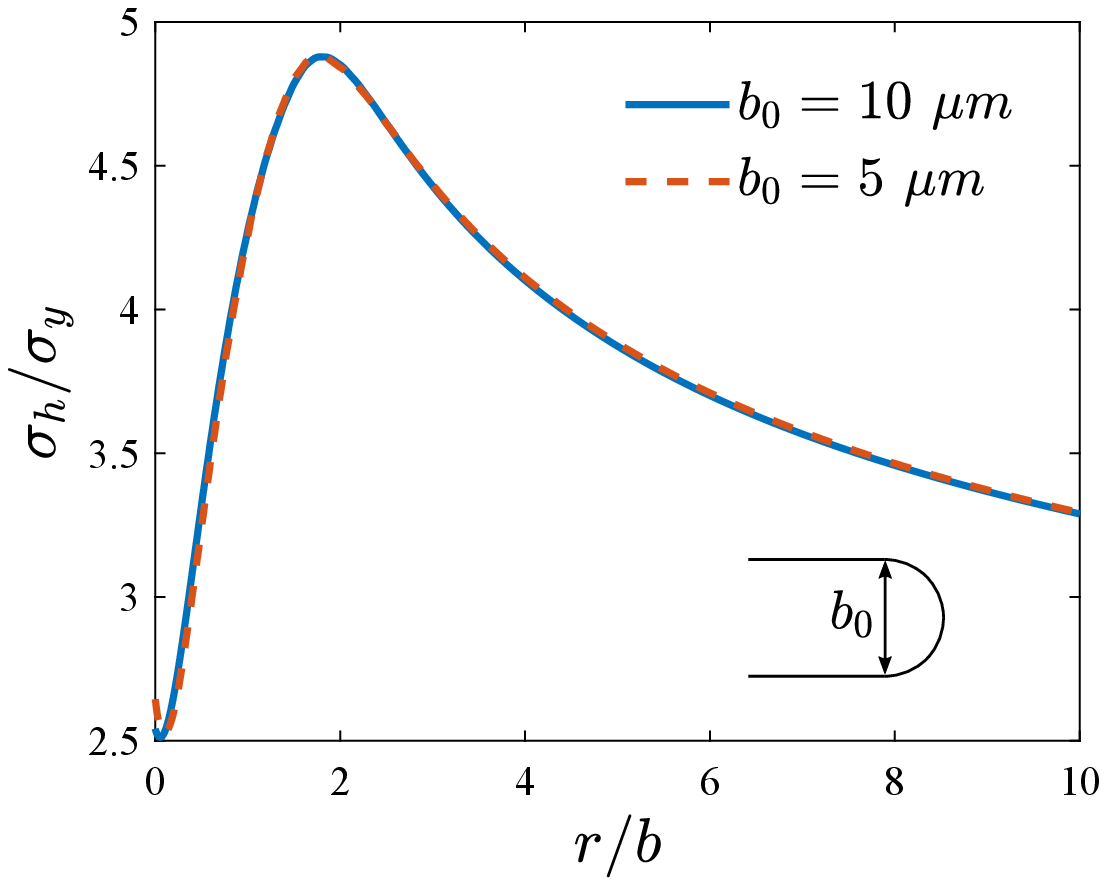}
                \caption{}
                \label{fig:SofMcMeekingSH}
        \end{subfigure}
        \begin{subfigure}[b]{0.53\textwidth}
                \raggedleft
                \includegraphics[scale=0.6]{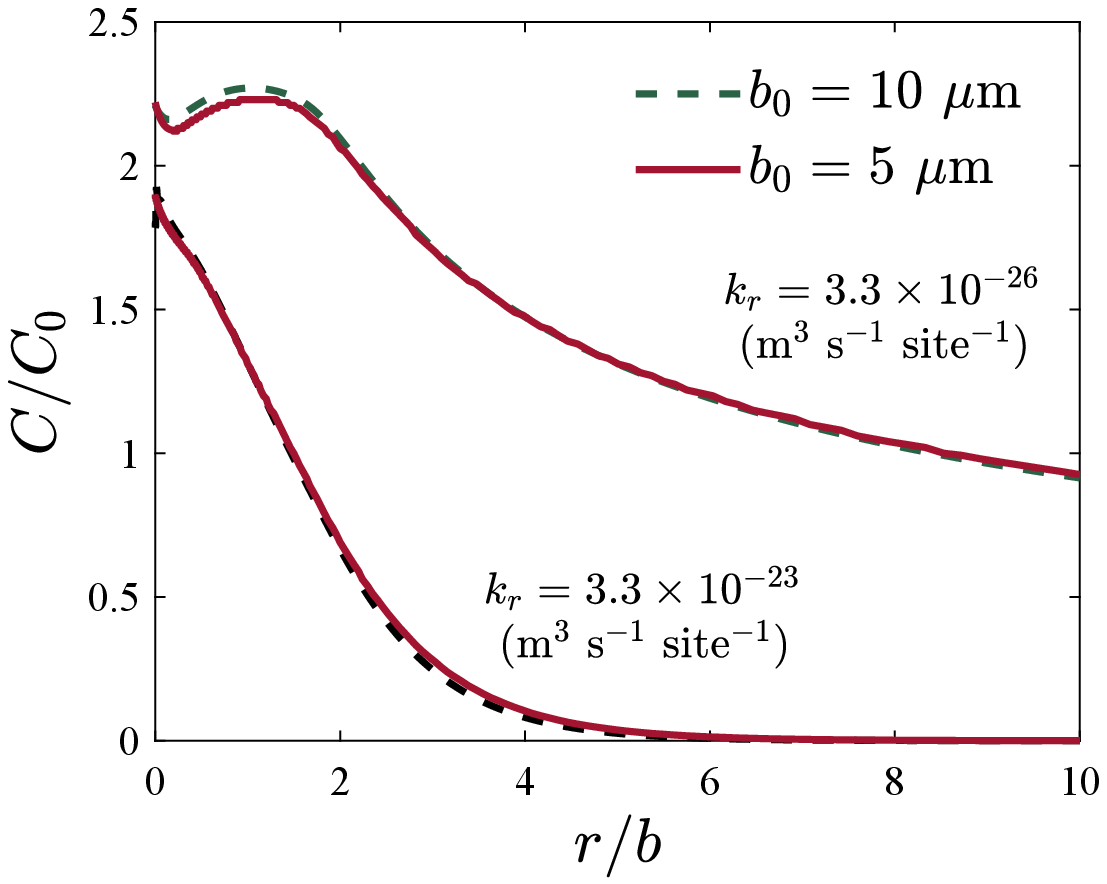}
                \caption{}
                \label{fig:SofMcMeekingC}
        \end{subfigure}}
        \caption{Influence of the crack tip opening. Distributions of (a) hydrostatic stress, and (b) lattice hydrogen concentration ahead of the crack for a remote load of $K_I=100$ MPa$\sqrt{m}$, a time of $t=130$ s and the material properties outlined in Tables \ref{Tab:diffcase2} and \ref{Tab:mechcase2}.} \label{fig:SofronisMcMeekingb0}
\end{figure}

We proceed to evaluate the role of the crack tip opening. Following the work by Sofronis and McMeeking \cite{Sofronis1989}, the above results have been computed for a specific choice of the initial crack tip blunting radius. Since the earlier work by McMeeking \cite{McMeeking1977a}, it is known that the hydrostatic stress distribution is independent of the initial crack tip blunting $b_0$ if the distance ahead of the crack is normalised by the current crack tip blunting $b$, and if the load is sufficiently large such that $b$ is at least five times larger than $b_0$. As we show in Fig. \ref{fig:SofronisMcMeekingb0}a, this can be accomplished by loads on the order of $K_I=100$ MPa$\sqrt{m}$ in iron-based materials with low yield stress ($\sigma_y=250$ MPa). Accordingly, the sensitivity of the hydrogen concentration to the crack tip opening is negligible under those conditions, see Fig. \ref{fig:SofronisMcMeekingb0}b. However, material systems of interest from an environmentally assisted cracking perspective often have a larger yield strength than 250 MPa and exhibit fracture at remote loads well below $K_I=100$ MPa$\sqrt{m}$. Moreover, cracks tips are significantly sharper in materials undergoing stress corrosion cracking or hydrogen embrittlement (see, e.g., Ref. \cite{Turnbull2017} and references therein). We explore more realistic conditions by extending Section \ref{Sec:AISI4340steel} to selected values of the initial crack tip blunting $b_0$. The hypothesis that the crack tip opening plays a fundamental role must be assessed for an accurate estimation of the hydrogen concentration distribution.

\begin{figure}[H]
        \begin{subfigure}[h]{1.1\textwidth}
                \centering
                \includegraphics[scale=0.85]{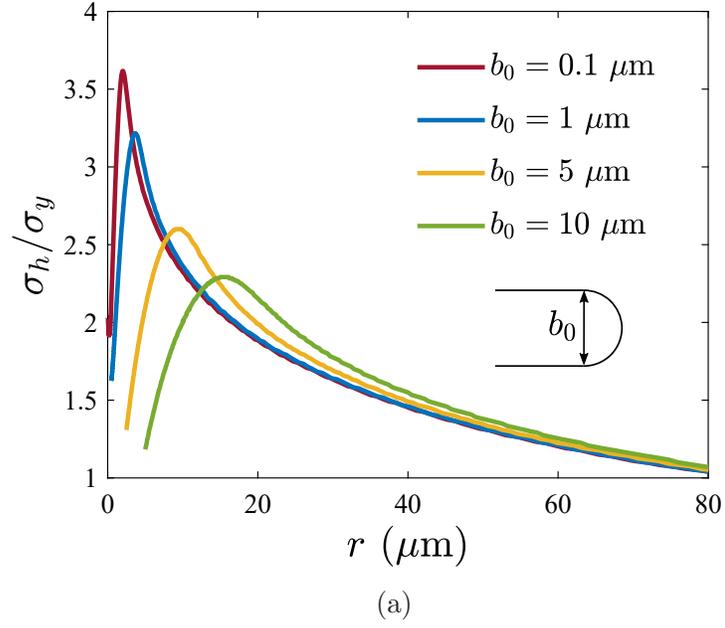}
                \caption{}
                \label{fig:Turnbullb0SH}
        \end{subfigure}\\
		
        \begin{subfigure}[h]{1.1\textwidth}
                \centering
                \includegraphics[scale=0.85]{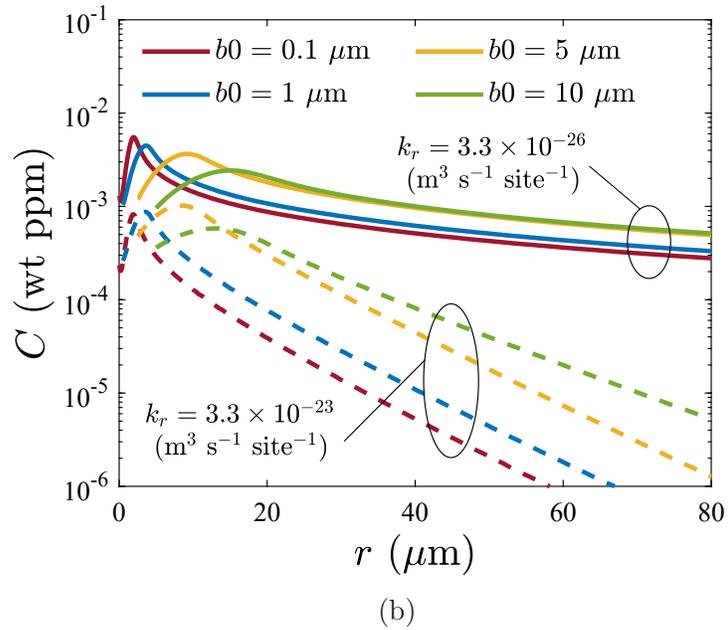}
                \caption{}
                \label{fig:Turnbullb0C}
        \end{subfigure}       
        \caption{Influence of the crack tip opening. Distributions of (a) hydrostatic stress, and (b) lattice hydrogen concentration ahead of the crack for selected values of the initial crack tip radius, a remote load of $K_I=30$ MPa$\sqrt{m}$, a time of $t=67$ s and the material properties outlined in Tables \ref{Tab:case0} to \ref{Tab:mechcase1}.}\label{fig:Turnbullb0}
\end{figure}

The results computed for $K_I=30$ MPa$\sqrt{m}$ and $t=67$ s using generalised boundary conditions are shown in Fig. \ref{fig:Turnbullb0} for selected values of the trapping rate constant $k_r$. We aim at gaining insight into critical distances in hydrogen assisted cracking and consequently show results along the extended crack plane $r$, without normalizing by $b$. Four values of the initial blunting are considered, covering the range $b_0=0.1$ $\mu$m to $b_0=10$ $\mu$m.\\

Consider first the hydrostatic stress results, Fig. \ref{fig:Turnbullb0SH}. Our calculations reveal that: (i) the maximum value of $\sigma_h$ attained increases with diminishing crack tip radius, and (ii) the location of the peak stress is closer to the crack tip for sharper cracks. The implications on the diffusion results are evident, see Fig. \ref{fig:Turnbullb0C}. For $k_r=3.3 \times 10^{-26}$ (m$^3$ s$^{-1}$ site$^{-1}$) the trends replicate those observed for $\sigma_h$: with diminishing $b_0$, the peak concentration increases and approaches the crack tip. The maximum concentration level is also closer to the crack tip for smaller values of $b_0$ when $k_r=3.3 \times 10^{-23}$ (m$^3$ s$^{-1}$ site$^{-1}$). However, the maximum value appears to be rather insensitive to changes in the initial crack tip blunting for $b_0 \leq 5$ $\mu$m.

\subsection{Influence of the trap density}
\label{Sec:InfluenceNr}

\begin{figure}[H]
        \begin{subfigure}[h]{1.1\textwidth}
                \centering
                \includegraphics[scale=0.85]{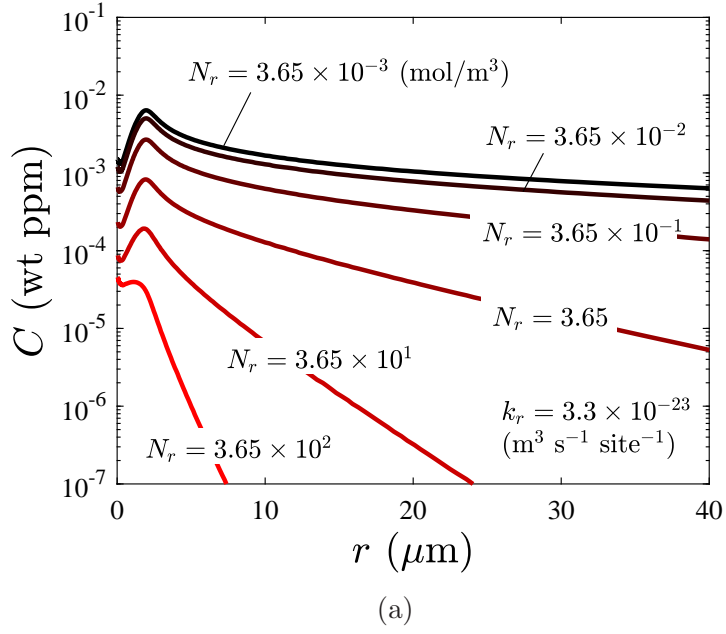}
                \caption{}
                \label{fig:TurnbullNrkrsmall}
        \end{subfigure}\\
		
        \begin{subfigure}[h]{1.1\textwidth}
                \centering
                \includegraphics[scale=0.85]{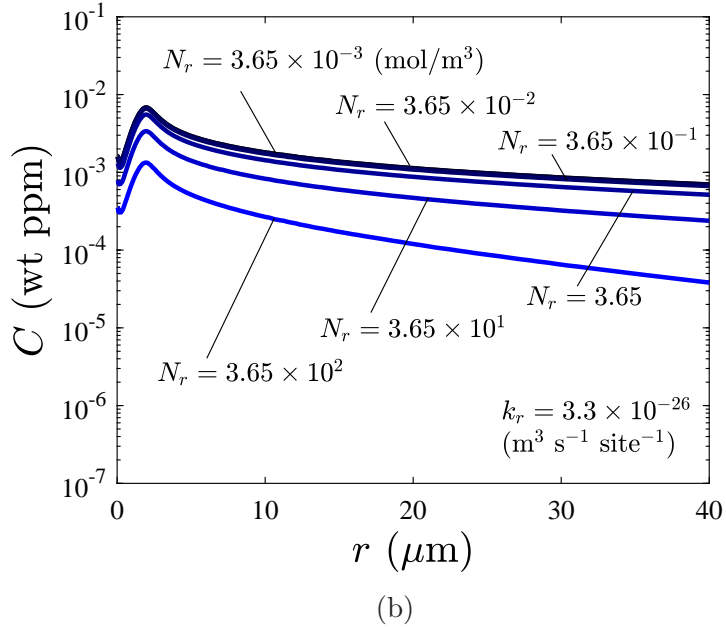}
                \caption{}
                \label{fig:TurnbullNrkrbig}
        \end{subfigure}       
        \caption{Influence of the trap density. Distributions of lattice hydrogen concentration for trapping rate constants (a) $k_r$: $3.3 \times 10^{-23}$ (m$^3$ s$^{-1}$ site$^{-1}$), and (b) $3.3 \times 10^{-26}$ (m$^3$ s$^{-1}$ site$^{-1}$). Remote load $K_I=30$ MPa$\sqrt{m}$, time $t=67$ s, and the material properties outlined in Tables \ref{Tab:case0} to \ref{Tab:mechcase1}.}\label{fig:TurnbullNr}
\end{figure}

The results presented for the first case study on AISI 4340 steel have been obtained with a trap density of $N_r=3.65$ mol/m$^3$, following Ref. \cite{Turnbull1996}. However, the specific value of $N_r$ is uncertain, as it depends on the type of trap. We extend the analysis of Section \ref{Sec:AISI4340steel} to compute the hydrogen concentration ahead of the crack for selected values of $N_r$. Results are shown in Figs. \ref{fig:TurnbullNrkrsmall} and \ref{fig:TurnbullNrkrbig} for two choices of $k_r$: $3.3 \times 10^{-23}$ m$^3$ s$^{-1}$ site$^{-1}$ and $3.3 \times 10^{-26}$ m$^3$ s$^{-1}$ site$^{-1}$, respectively. \\

In agreement with expectations, the influence is significantly higher for a high trapping rate constant, $k_r= 3.3 \times 10^{-23}$ (m$^3$ s$^{-1}$ site$^{-1}$). Qualitatively, the trend is the same in Figs. \ref{fig:TurnbullNrkrsmall} and \ref{fig:TurnbullNrkrbig}; the larger the trap density the lower the hydrogen concentration in lattice sites. Note that the crack tip hydrogen concentration, i.e. $C$ for $r = 0$, is sensitive to $N_r$ in the context of generalised boundary conditions, as opposed to the conventional constant hydrogen concentration boundary conditions. For low values of $k_r$, results show differences of several orders of magnitude for the range of $N_r$ values considered.

\subsection{Strain gradient plasticity}
\label{Sec:StrainGradientPlasticity}

Plasticity and dislocation density can have a profound effect on crack tip hydrogen concentration. For example, Lekbir \textit{et al.} \cite{Lekbir2013} investigated the influence of dislocation density on the number of potential adsorption/desorption sites as well as in the activation energies involved in the HER. Experimentally, plastic straining has been demonstrated to increase cathodic current densities on nickel \cite{ElAlami2006}. Of interest here is the influence of crack tip dislocation hardening mechanisms in elevating the stresses. Plastic strain gradients are associated with lattice curvature and geometrically necessary dislocations (GNDs) \cite{Ashby1970}, and the resulting increased dislocation density promotes strengthening. Flow stress elevation in the presence of plastic strain gradients has been measured in a wide range of mechanical tests on micro-sized samples, such as indentation \cite{Nix1998}, torsion \cite{Fleck1994}, and bending \cite{Stolken1998}. These experiments show a three-fold increase in the effective flow stress by reducing the size of the specimen (\emph{smaller is stronger}). Strain gradient plasticity theory has been developed to capture these dislocation hardening mechanisms \cite{Gao1999,Fleck2001,Gudmundson2004,Gurtin2005}. The plastic work is defined in terms of both the plastic strain and plastic strain gradient, introducing a length scale in the material description. Strain gradient hardening is expected to play a big role in fracture where, independently of the size of the specimen, the plastic zone adjacent to the crack tip is physically small and contains strong spatial gradients of deformation. The analysis of crack tip fields ahead of stationary or propagating cracks using strain gradient plasticity reveals a notable stress elevation relative to conventional plasticity predictions \cite{Wei1997,Komaragiri2008,IJSS2015,EJMAS2019}. This stress elevation can have an important effect in predicting hydrogen assisted cracking, given the exponential dependence on hydrogen concentration with hydrostatic stresses and the micro-scale critical distance for cracking \cite{Gangloff2003a}.\\

We investigate the role of plastic strain gradients in altering the hydrostatic stress concentration by coupling the present hydrogen transport framework to the Gudmundson \cite{Gudmundson2004} higher order strain gradient plasticity model. Strain gradient effects are accounted for \textit{via} the free energy and the definition of a gradient-enhanced equivalent plastic strain. The former is given as a function of elastic strains $\varepsilon_{ij}^e$ and plastic strain gradients $\varepsilon_{ij,k}^p$ as,
\begin{equation}
    \Psi \left( \varepsilon_{ij}^e , \, \varepsilon_{ij,k}^p \right) = \frac{1}{2} \varepsilon_{ij}^e C_{ijkl} \varepsilon_{kl}^e + \frac{1}{2} \mu L_E^2 \varepsilon_{ij,k}^p \varepsilon_{ij,k}^p
\end{equation}

\noindent where $C_{ijkl}$ is the isotropic elastic stiffness tensor, $\mu$ is the shear modulus and $L_E$ is the so-called energetic material length scale. On the other side, the generalised effective plastic strain rate $\dot{E}^p$ reads:
\begin{equation}
    \dot{E}^p = \left( \frac{2}{3} \dot{\varepsilon}_{ij}^p \dot{\varepsilon}_{ij}^p + L_D^2 \dot{\varepsilon}_{ij,k}^p \dot{\varepsilon}_{ij,k}^p \right)^{1/2}
\end{equation}

\noindent where $L_D$ is a dissipative material length scale. Modern strain gradient plasticity theories include both energetic and dissipative length scales to capture the hardening and strengthening behaviours observed in the experiments. For simplicity, we choose to define a reference scale $\ell=L_E=L_D$, with the conventional plasticity case recovered when $\ell=0$. The numerical implementation is given in Ref. \cite{JMPS2019} and will not be described here for the sake of brevity.\\

\begin{figure}[H]
  \makebox[\textwidth][c]{\includegraphics[width=1\textwidth]{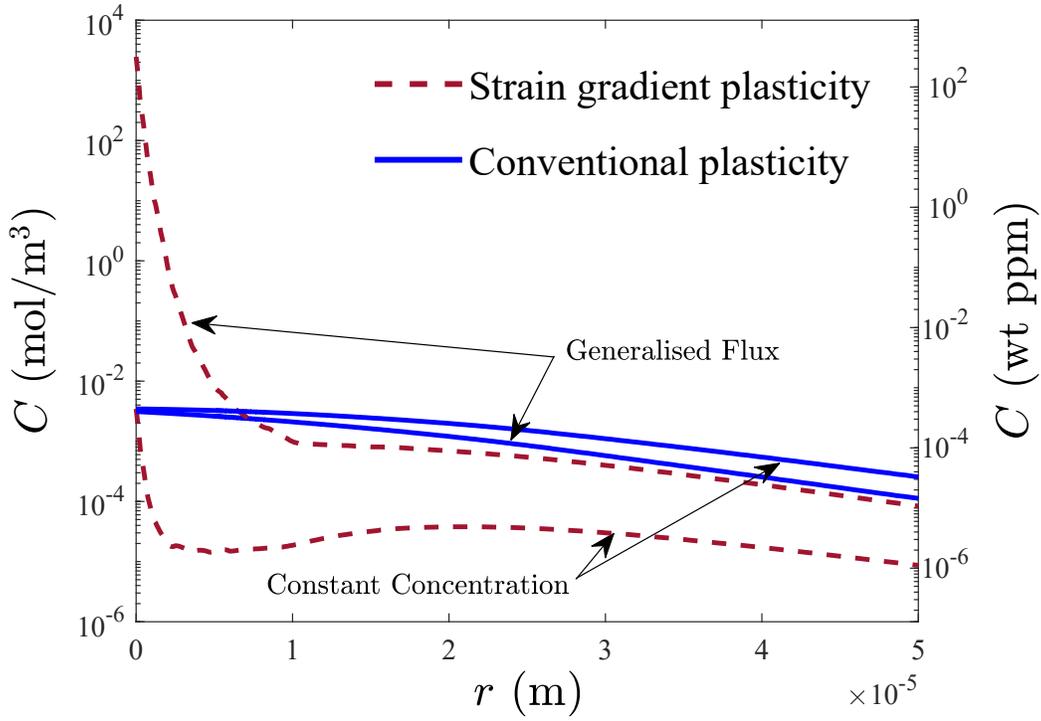}}%
  \caption{Dislocation hardening effects. Lattice hydrogen distribution ahead of the crack tip predicted by strain gradient plasticity ($\ell=5$ $\mu$m) and conventional plasticity for a remote load of $K_I=30$ MPa$\sqrt{m}$, time $t=130$ s and $k_r=3.4 \times 10^{-23}$ m$^3$/(site s). Iron-based model material with properties described in Tables \ref{Tab:diffcase2} and \ref{Tab:mechcase2}.}
  \label{fig:FigSGPCCvsGF}
\end{figure}

Crack tip hydrogen distributions are computed for the second case study, the iron-based model material addressed by Sofronis and McMeeking \cite{Sofronis1989}. Material properties and initial hydrogen concentration are those given in Tables \ref{Tab:diffcase2} and \ref{Tab:mechcase2}. The material length scale associated with plastic strain gradients is assumed to be equal to $\ell=5$ $\mu$m, an intermediate value within the range of length scales reported in the literature from micro-scale experiments \cite{IJP2020}. Unlike the analysis of Section \ref{Sec:SofronisMcMeeking}, the trap density is assumed to be constant and equal to $N_r=2.2 \times 10^{24}$ sites/m$^3$. Results are shown in Fig. \ref{fig:FigSGPCCvsGF} for a remote load of $K_I=30$ MPa$\sqrt{m}$, a total time of $t=130$ s and a trapping rate constant of $k_r=3.4 \times 10^{-23}$ m$^3$/(site s). For the sake of clarity, the vertical axis is shown in logarithmic scale.

The results reveal interesting features. First, for the strain gradient plasticity case, differences of up to six orders of magnitude in the crack tip hydrogen concentration are predicted when considering generalised flux versus constant concentration boundary conditions. The hydrostatic stress raises sharply as in the vicinity of the crack but the constraint of a constant concentration at the crack faces reduces the hydrogen distribution even beyond the conventional plasticity assumption. In other words, the use of constant concentration schemes is not suitable for gradient-enhanced models. When considering the generalised flux predictions, strain gradient plasticity predicts a crack tip hydrogen concentration that is much larger than the conventional plasticity result. Such high hydrogen concentrations close to the crack tip agree with neutron activation measurements \cite{Gerberich2012}, and rationalise decohesion-based arguments \cite{JMPS2020}. High crack tip concentrations have also been reported in SIMS analyses that do not distinguish between lattice and trapped hydrogen concentration \cite{Mao1998}.\\

\begin{figure}[H]
  \makebox[\textwidth][c]{\includegraphics[width=1\textwidth]{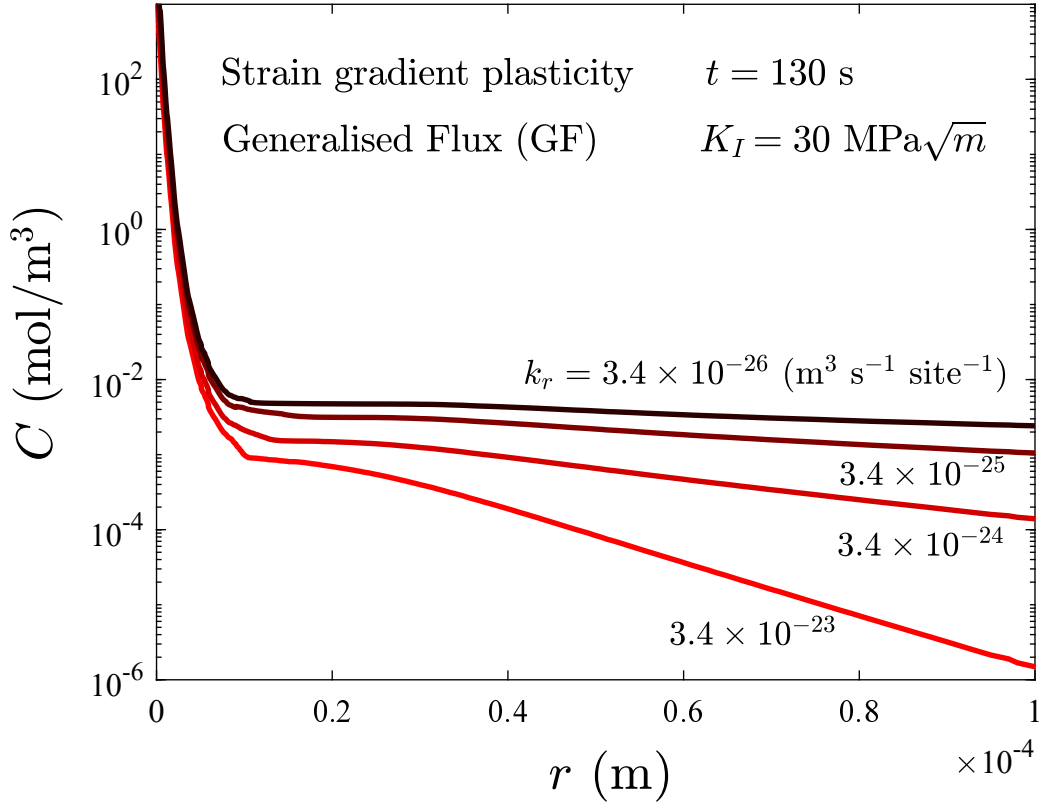}}%
  \caption{Dislocation hardening effects. Hydrogen distribution ahead of the crack tip predicted by strain gradient plasticity ($\ell=5$ $\mu$m) for different $k_r$ values, a remote load of $K_I=30$ MPa$\sqrt{m}$ and a total time $t=130$ s. Iron-based model material with properties described in Tables \ref{Tab:diffcase2} and \ref{Tab:mechcase2}.}
  \label{fig:FigSGPkr}
\end{figure}

Finally, we assess the role of the trapping rate constant $k_r$ in the lattice hydrogen distribution predicted by strain gradient plasticity. The results are shown in Fig. \ref{fig:FigSGPkr} for the same conditions as the previous figure but selected values of $k_r$. The qualitative trends follow those observed in conventional plasticity (see, e.g., Fig. \ref{fig:Fig2K30GF_kr}), with the hydrogen concentration increasing with decreasing $k_r$. In all cases, the hydrogen concentration raises sharply at approximately 10 $\mu$m from the crack tip.

Further insight into the role of plastic deformation across scales can be obtained combining the present generalised boundary conditions with conventional crystal plasticity or strain gradient crystal plasticity \cite{Pouillier2012,Jothi2015,Charles2017}. The influence of other effects, such as texture, can be characterised provided that the diffusion and adsorption/absorption constants are adequately measured considering material anisotropy \cite{Li2017}.


\section{Conclusions}
\label{Concluding remarks}

We present a generalised framework for modelling hydrogen transport at crack tips. The model combines, for the first time, (i) McNabb-Foster trapping kinetics, (ii) generalised boundary conditions to capture the absorption/adsorption fluxes, and (iii) finite strain plasticity, as given by J2 flow theory or strain gradient plasticity. These features enable capturing the hydrostatic stress dependence of surface concentration and extend the applicability of hydrogen diffusion simulations beyond the range of scenarios where the equilibrium assumption is appropriate. The generalised framework presented is implemented in a finite element setting, rigorously validated, and used to gain insight into trapping and surface phenomena. Model predictions are showcased by addressing two material systems: a high-strength alloy (AISI 4340 steel) and the model iron-based material used in the paradigmatic study by Sofronis and McMeeking \cite{Sofronis1989}. The impact on modelling predictions of using generalised boundary conditions is demonstrated. Absorption/adsorption constants, that should be experimentally determined for different material and electrolyte conditions, influence hydrogen uptake and the magnitude of hydrogen lattice concentration near a crack tip. The role of trap density and crack radius is also assessed. Moreover, since hydrostatic stress is an important variable in deviating hydrogen transport from ideal diffusion, the influence of strain gradient plasticity is also assessed, so as to provide a richer description of crack tip fields and hydrogen accumulation. Our main findings are:
\begin{itemize}
  \item The use of generalised flux boundary conditions leads to crack tip hydrogen concentrations that can be several orders of magnitude larger than those predicted by the common constant hydrogen concentration approach. Differences due to surface kinetics, the effect of the hydrostatic stress and trap density are quantified.
  \item The initial crack tip blunting plays an important role in quantifying the hydrogen concentration for remote loads and material properties relevant to hydrogen embrittlement.
  \item Constant concentration boundary conditions fail to capture the enhancement in hydrogen concentration associated with dislocation hardening. The coupling of generalised boundary conditions and strain gradient plasticity reveals very high hydrogen concentrations close to the crack surface.
\end{itemize}
 

\section{Acknowledgments}
\label{Acknowledge of funding}

E. Mart\'{\i}nez-Pa\~neda acknowledges financial support from EPSRC funding under grant No. EP/R010161/1, from the UKCRIC Coordination Node EPSRC grant number EP/R017727/1, which funds UKCRIC's ongoing coordination, and from Wolfson College Cambridge (Junior Research Fellowship). A. D\'{\i}az gratefully acknowledges financial support from the Ministry of Science, Innovation and Universities of Spain through grant RTI2018-096070-B-C33.

\section{Data availability statement}
\label{Sec:DataStatemenet}

The data generated during this study will be made available upon reasonable request.


\appendix
\section{Numerical verification - McNabb and Foster (TDS)}
\label{App:TDS}

The framework presented here constitutes the first finite element implementation of a coupled mechanical-diffusion model based on McNabb-Foster kinetics and including generalised boundary conditions. Accordingly, validation of the numerical implementation is done as a three-stage process. First, we show that our McNabb-Foster diffusion model reproduces the results by Legrand \emph{et al.} \cite{Legrand2015} in modelling thermal desorption spectroscopy (TDS).\\

By assuming only radial diffusion, for a specimen of radius $a$, the problem becomes one dimensional. Hydrogen transport is modelled with Eq. (\ref{Eq:Htransport}), without the mechanical coupling ($\sigma_h=0$). We define the diffusion coefficient as,
\begin{equation}
    D=D_0 \exp \left[ - \frac{E_L}{R \left( \phi t + T_i \right)} \right]
\end{equation}

\noindent where $E_L$ is the activation energy for lattice diffusion, $D_0$ is the pre-exponential factor for the lattice diffusion coefficient, $\phi$ (K s$^{-1}$) is the temperature ramp and $T_i$ is the initial temperature. And we define the rate constants as,
\begin{equation}
    k_r=k_r^0 \exp \left[ - \frac{E^t}{R \left( \phi t + T_i \right)} \right]
\end{equation}
\begin{equation}
    p_r=p_r^0 \exp \left[ - \frac{E^d}{R \left( \phi t + T_i \right)} \right]
\end{equation}

\noindent Here, $k_r^0$ and $p_r^0$ are the pre-exponential constants for the capture and release rates, and $E^t$ and $E^d$ are the activation energies for capture (trapping) and release (detrapping). The parameters adopted in this analysis are shown in Table \ref{Tab:TDSparameters}, following Ref. \cite{Legrand2015}. The difference between detrapping and trapping energies represents, by definition, the binding energy; Legrand \emph{et al.} \cite{Legrand2015} choose to simulate a trap with $E_B$ = 44.4 kJ/mol. \\

\begin{table}[H]
\centering
\caption{TDS model parameters, following Ref. \cite{Legrand2015}.}
\label{Tab:TDSparameters}
   {\tabulinesep=1.2mm
   \makebox[\textwidth][c]{\begin{tabu} {cccccccccccc}
       \hline
 $D_0$ & $N_r$ & $E_L ; E^t$ & $k_r^0$ & $E^d$ & $p_r^0$ \\ \hline
 $2.74 \times$10$^{-6}$ & $1.2 \times$10$^{24}$ & 19.29 & $4.74 \times$10$^{7}$ & 53.69 & $1.0 \times$10$^{8}$  \\
 (m$^2$/s) & (sites/m$^{3}$) & (kJ/mol) & (m$^3$/mol s) & (kJ/mol)   &  (s$^{-1}$) \\\hline
   \end{tabu}}}
   {\tabulinesep=1.2mm
   \makebox[\textwidth][c]{\begin{tabu} {cccccccccccc}

 $a$   &  $T_i$ & $\phi$ & $C_0$ & $\theta_{r,0}$ \\ \hline
 2 & 10 & 50 &  1.0 &   1.0 \\
 (mm) & (K) & (K/min) & (mol/m$^{3}$) & (-)\\\hline
   \end{tabu}}}
\end{table}

Equations (\ref{Eq:Htransport}) and (\ref{Eq:thetaR}) are solved by defining the following initial and boundary conditions. First, the specimen is assumed to be charged uniformly: $C=C_0$ at $t=0$ for all $x$; traps are considered to be completely filled at this initial time due to the high binding energy, $\theta_{r,0} = 1.0$. At time greater than zero we assume that the concentration of hydrogen at the surface is zero: $C=0$ at $t>0$ at $x=0$. In addition, we take advantage of symmetry and model half of the slab, prescribing a zero flux at the mid point: $\partial C / \partial x=0$ at $x=a$.\\

\begin{figure}[H]
  \makebox[\textwidth][c]{\includegraphics[width=1\textwidth]{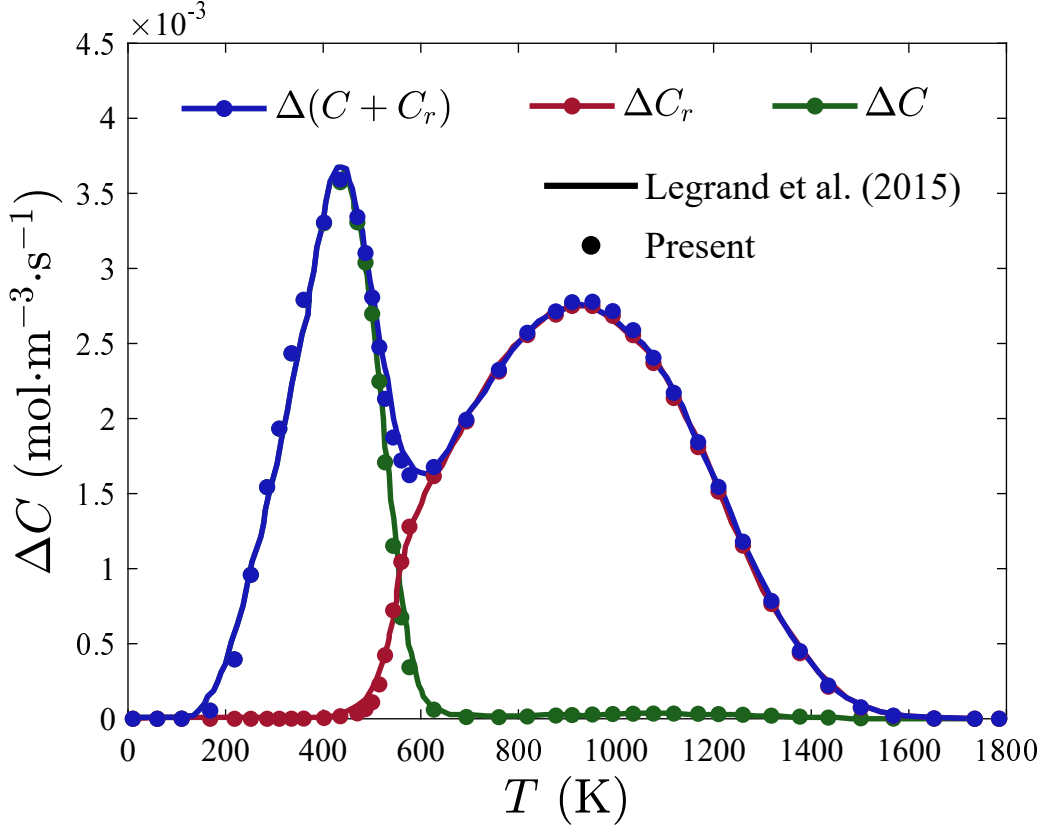}}%
  \caption{TDS desorption spectrum predictions for the lattice and trapped hydrogen. Comparison of present results (symbols) with Legrand \emph{et al.} \cite{Legrand2015} (digitalized lines).}
  \label{fig:Calibration}
\end{figure}

The results obtained are shown in Fig. \ref{fig:Calibration} in terms of the quantity of
hydrogen that escaped the simulated TDS specimen at each time. Both for the lattice sites and the traps, $\Delta C$ is computed by integrating the hydrogen concentration over the slab length and dividing it by the time increment. The solid curve $\Delta (C + C_r)$ represents the desorption of the total hydrogen concentration. Results show a very good agreement with the work by Legrand \emph{et al.} \cite{Legrand2015}.

\section{Numerical verification - Electrochemical Permeation}
\label{App:EP}

The second step in validating the model involves verifying the implementation of the Neumann-type generalised boundary conditions. This is achieved by reproducing the modelling of electrochemical permeation tests conducted by Turnbull and co-workers \cite{Turnbull2014,Turnbull2015}. The relevant material parameters are listed in Tables \ref{Tab:EP_verif_Diff} and \ref{Tab:EP_verif_GF}. The charging constant $k_c$ can be expressed in units of an equivalent input current density through Faraday's constant, such that $5\times$10$^{-6}$ mol/(s$\cdot$m$^2$) is equivalent to 0.48 A/m$^2$.

\begin{table}[H]
\centering
\caption{Diffusion, mechanical and geometrical parameters for the verification study, following \cite{Turnbull2015}}
\label{Tab:EP_verif_Diff}
   {\tabulinesep=1.2mm
   \makebox[\textwidth][c]{\begin{tabu} {cccccccccccccc}
       \hline
$D_L$  & $N_r$ & $k_r$ & $p_r$\\ \hline
$7.2 \times$10$^{-9}$ & $2.2 \times$10$^{24}$ & $3.4 \times$10$^{-23}$ & 0.031 
\\
(m$^2$/s) & (sites/m$^3$) &  (m$^3$/(s$\cdot$site)) & (1/s)\\\hline   
   \end{tabu}}}
\end{table}

\begin{table}[H]
\centering
\caption{Parameters related to the boundary conditions for the permeation simulation, following \cite{Turnbull2015}}
\label{Tab:EP_verif_GF}
   {\tabulinesep=1.2mm
   \makebox[\textwidth][c]{\begin{tabu} {ccccc}
       \hline
$k_{abs}$  & $k_c$ & $k_{des}$ & $k_{r,chem}$ & $k_{r,elec}$\\ 
\hline
$1 \times$10$^{11}$ & $5 \times$10$^{-6}$ & $8.8 \times$10$^{9}$ & 22 & $5 \times$10$^{-3}$\\
(mol/(s$\cdot$m$^2$)) & (mol/(s$\cdot$m$^2$)) & (m/s) & (mol/(s$\cdot$m$^2$)) & (mol/(s$\cdot$m$^2$))\\\hline
   \end{tabu}}}
\end{table}

\begin{figure}[H]
        \begin{subfigure}[h]{1.1\textwidth}
                \centering
                \includegraphics[scale=0.7]{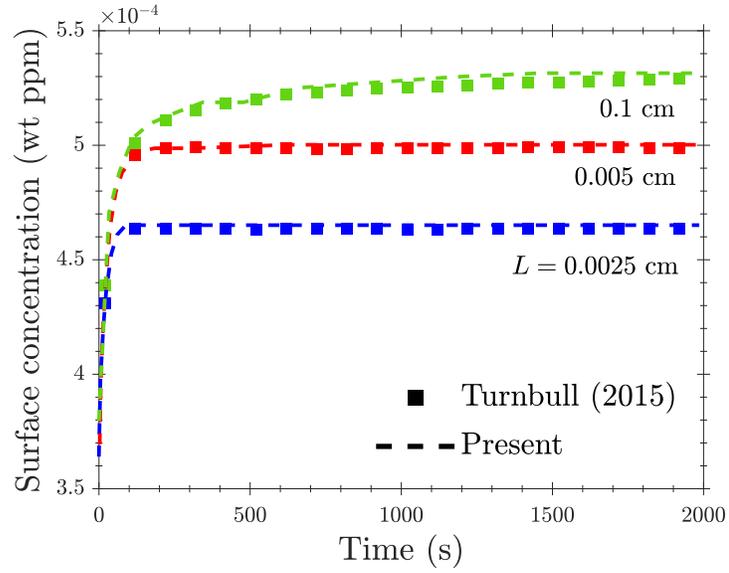}
                \caption{}
                \label{fig:Cs_verif}
        \end{subfigure}\\
		
        \begin{subfigure}[h]{1.1\textwidth}
                \centering
                \includegraphics[scale=0.7]{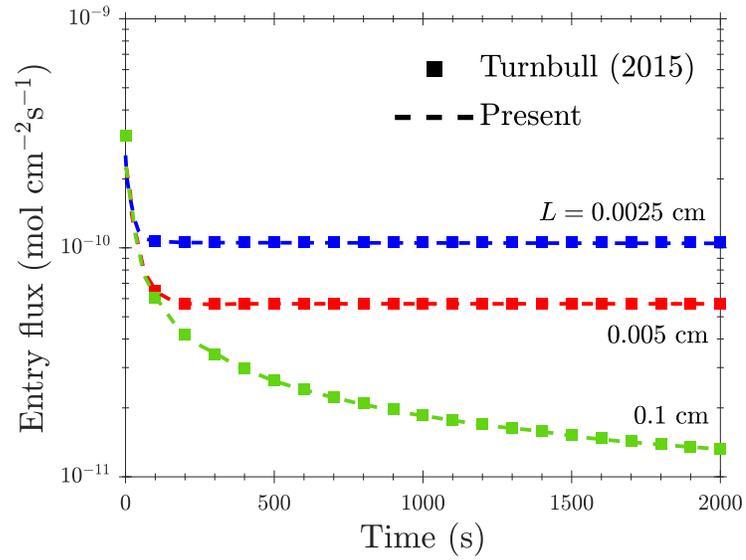}
                \caption{}
                \label{fig:Jin_verif}
        \end{subfigure}       
        \caption{Thickness effect with $L$ in cm on (a) sub-surface concentration and (b) hydrogen entry flux due to generalised boundary conditions and a charging constant of $k_c = 0.48$ $A/m^{2}$.}\label{fig:EPverification}
\end{figure}

The results obtained shown in Fig. \ref{fig:EPverification}, along with those obtained from Refs. \cite{Turnbull2014,Turnbull2015}. An excellent agreement is observed, quantitatively capturing the thickness effect on surface concentration and entry flux.

\section{Parametric study of entry constants}
\label{App:k_influence}

We aim at gaining insight into the role of the surface kinetics parameters entering the model. The evolution of surface variables, i.e. $C_{s}$ and $J_{in}$, is evaluated in this Appendix to predict hydrogen entry from a crack wall and a crack tip. However, results are shown for a permeation simulation in which an extremely thick specimen is reproduced ($L$ = 1 m). Thus, the effect of the exit surface is negligible and the evolution of hydrogen entry can be extrapolated to the crack surfaces. 
\\

\begin{figure}[H]
\makebox[\linewidth][c]{%
        \begin{subfigure}[b]{0.6\textwidth}
                \centering
                \includegraphics[scale=0.55]{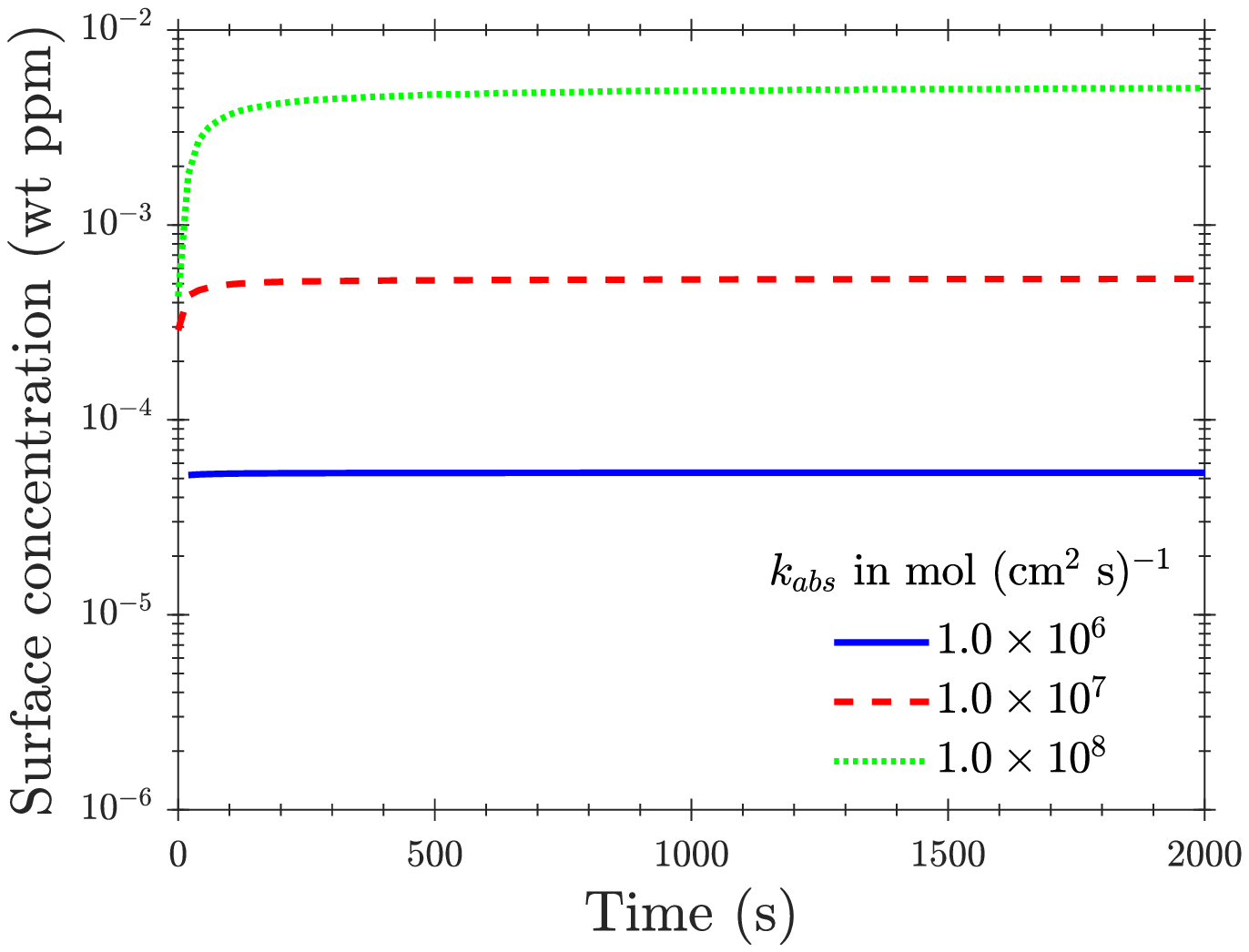}
                \caption{}
                \label{fig:Cs_kabs}
        \end{subfigure}
        \begin{subfigure}[b]{0.6\textwidth}
                \raggedleft
                \includegraphics[scale=0.55]{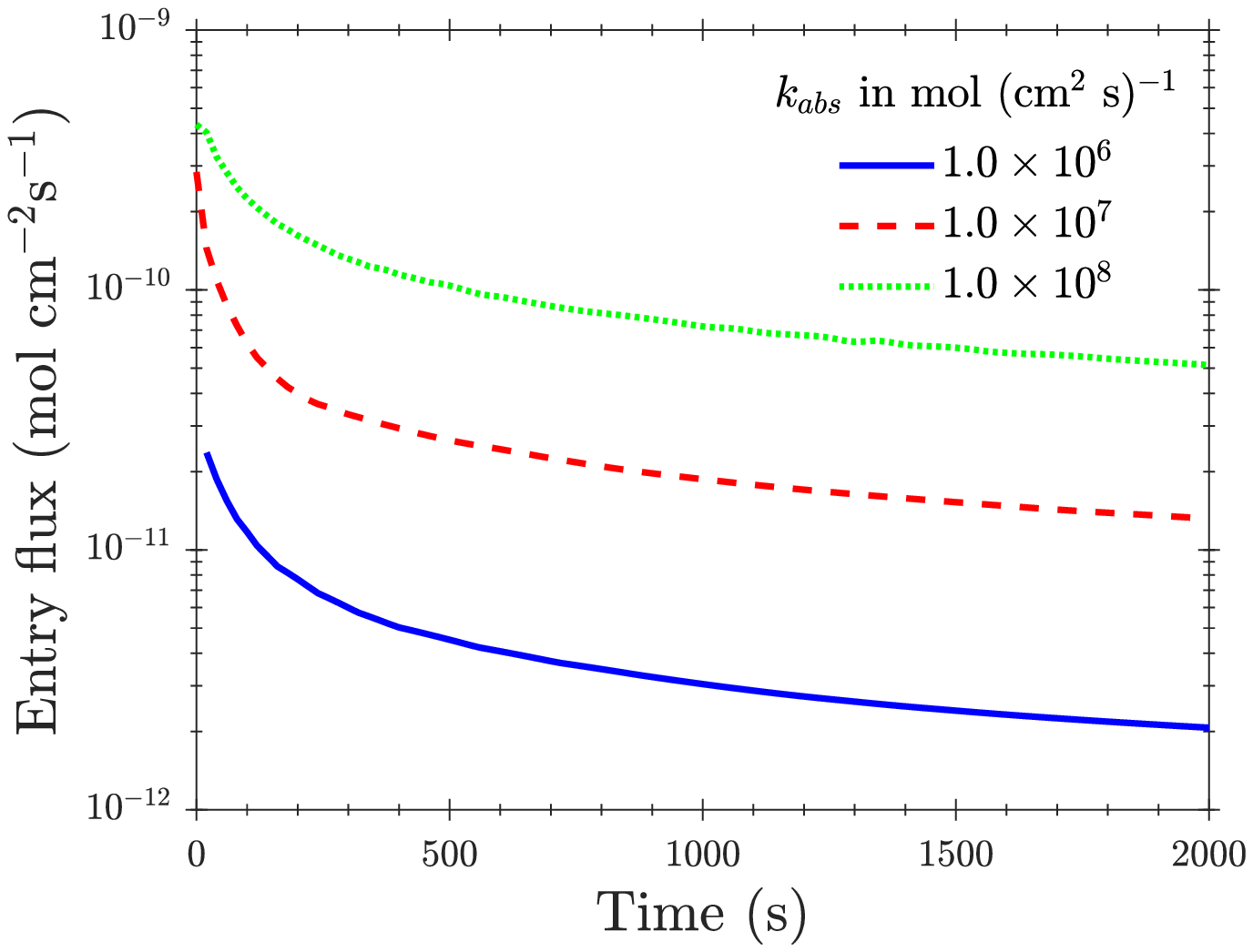}
                \caption{}
                \label{fig:Jin_kabs}
        \end{subfigure}}
        \caption{Influence of $k_{abs}$ on (a) sub-surface concentration and (b) hydrogen entry flux due to generalised boundary conditions}
        \label{fig:kabs}
\end{figure}

\begin{figure}[H]
\makebox[\linewidth][c]{%
        \begin{subfigure}[b]{0.6\textwidth}
                \centering
                \includegraphics[scale=0.55]{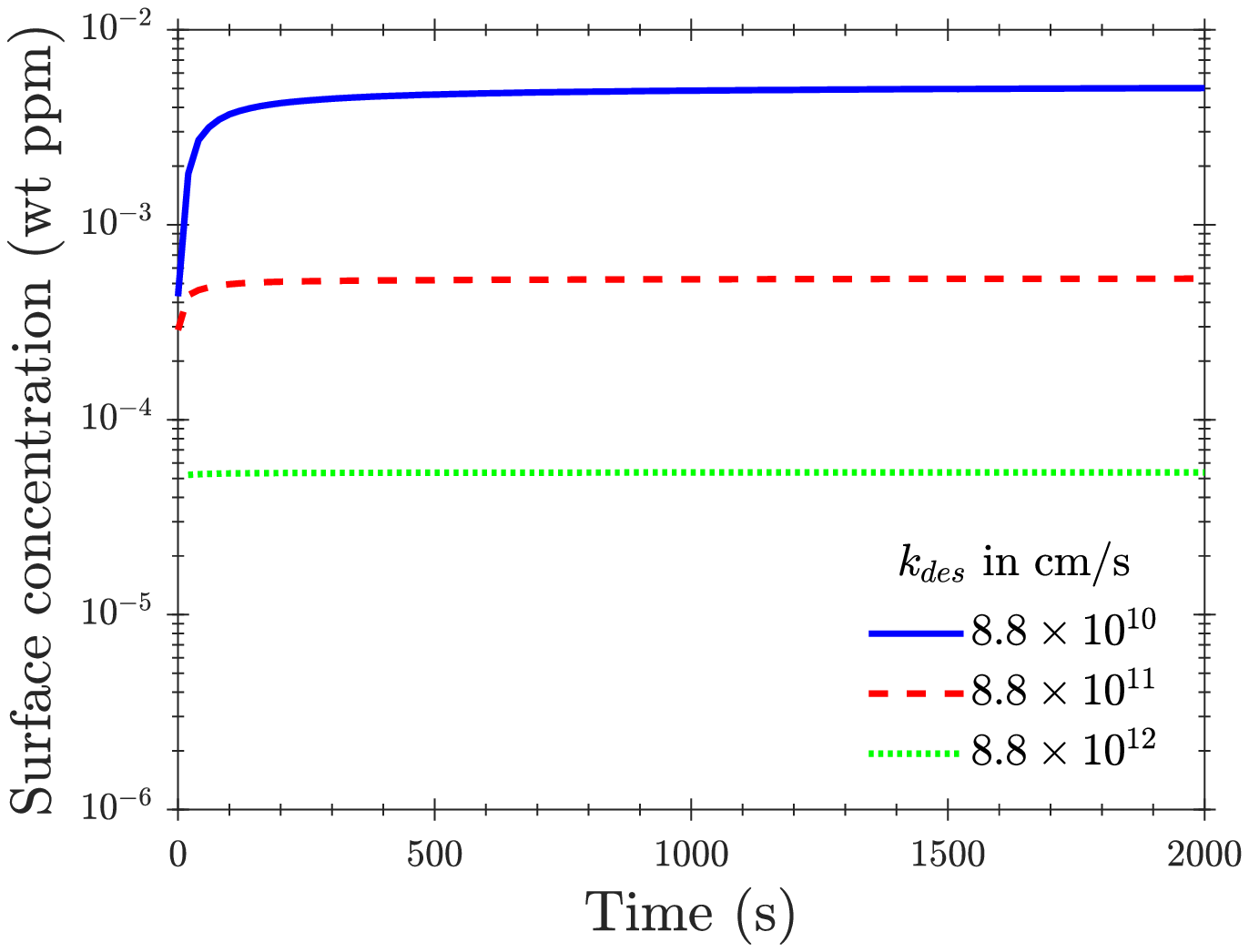}
                \caption{}
                \label{fig:Cs_kdes}
        \end{subfigure}
        \begin{subfigure}[b]{0.6\textwidth}
                \raggedleft
                \includegraphics[scale=0.55]{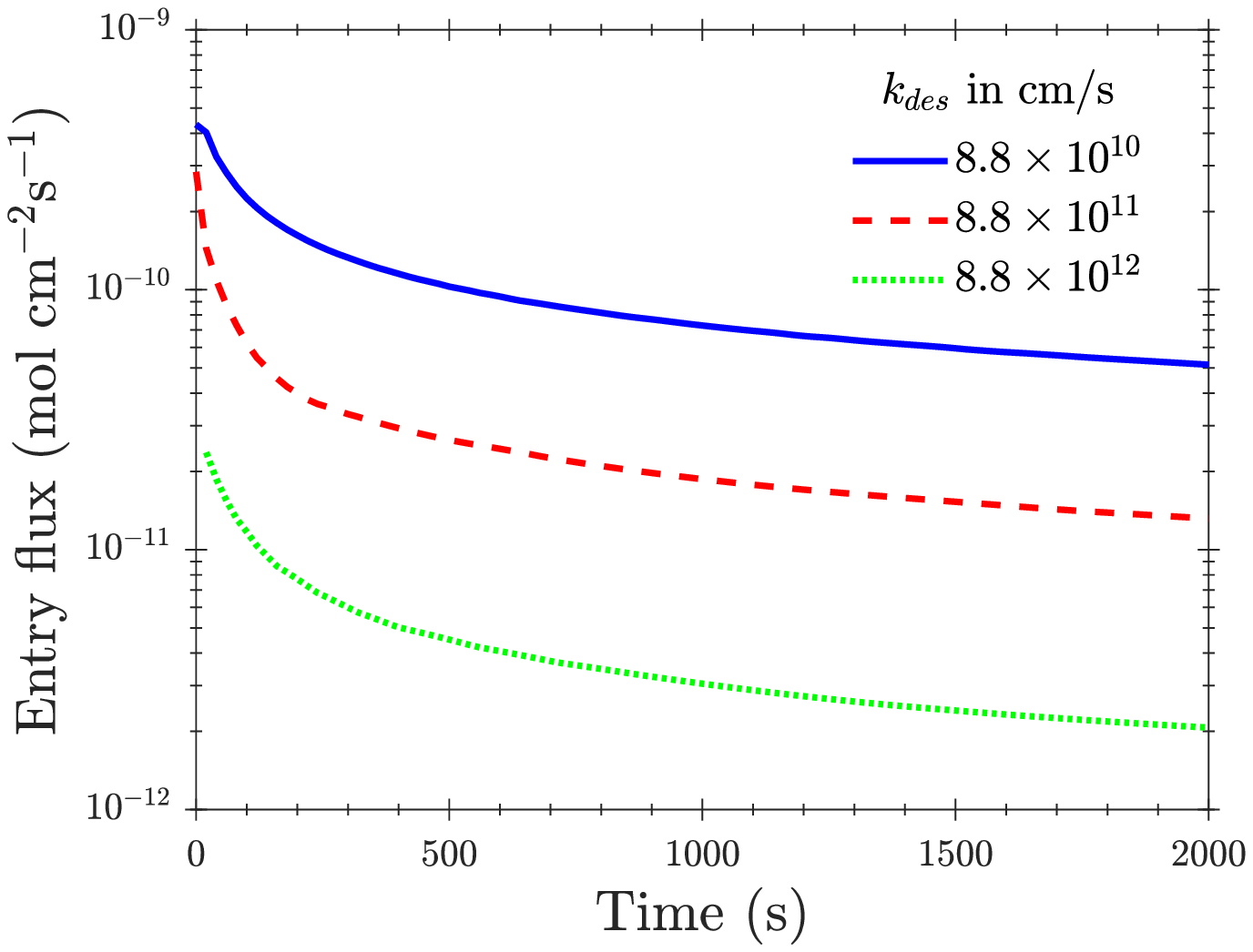}
                \caption{}
                \label{fig:Jin_kdes}
        \end{subfigure}}
        \caption{Influence of $k_{des}$ on (a) sub-surface concentration and (b) hydrogen entry flux due to generalised boundary conditions}
        \label{fig:kdes}
\end{figure}

The influence of absorption and desorption constants is assessed in Fig. (\ref{fig:kabs}) and Fig.  (\ref{fig:kdes}), respectively. As predicted by Eq. (\ref{Eq:surface_concentr}), the higher $k_{abs}$, the higher sub-surface concentration, whereas the opposite effect is found for $k_{des}$. The comparison between Figs. \ref{fig:kabs} and \ref{fig:kdes} reveals that the ratio $k_{abs}/k_{des}$ is the critical value that influences hydrogen uptake.

\begin{figure}[H]
\makebox[\linewidth][c]{%
        \begin{subfigure}[b]{0.6\textwidth}
                \centering
                \includegraphics[scale=0.55]{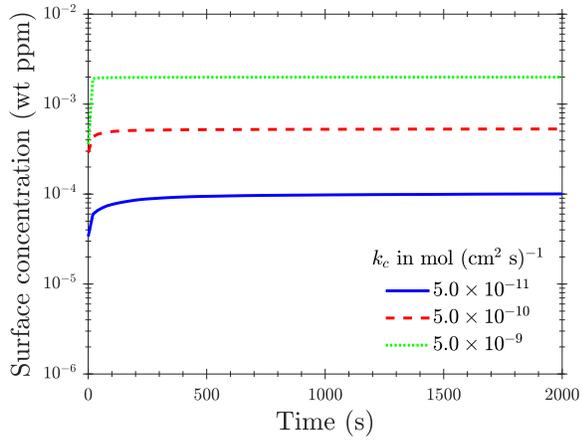}
                \caption{}
                \label{fig:Cs_kc}
        \end{subfigure}
        \begin{subfigure}[b]{0.6\textwidth}
                \raggedleft
                \includegraphics[scale=0.55]{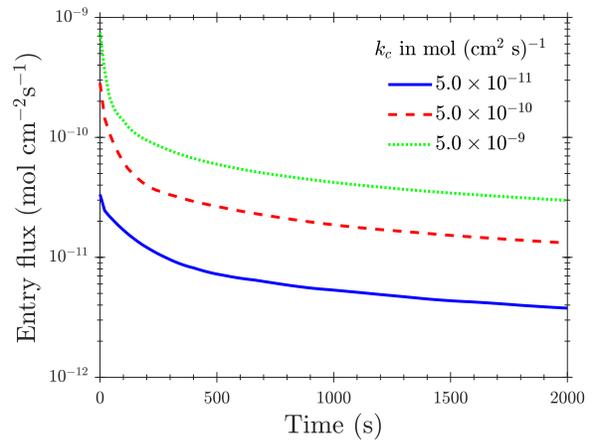}
                \caption{}
                \label{fig:Jin_kc}
        \end{subfigure}}
        \caption{Influence of $k_{c}$ on (a) sub-surface concentration and (b) hydrogen entry flux due to generalised boundary conditions}
        \label{fig:kc}
\end{figure}

\begin{figure}[H]
\makebox[\linewidth][c]{%
        \begin{subfigure}[b]{0.6\textwidth}
                \centering
                \includegraphics[scale=0.55]{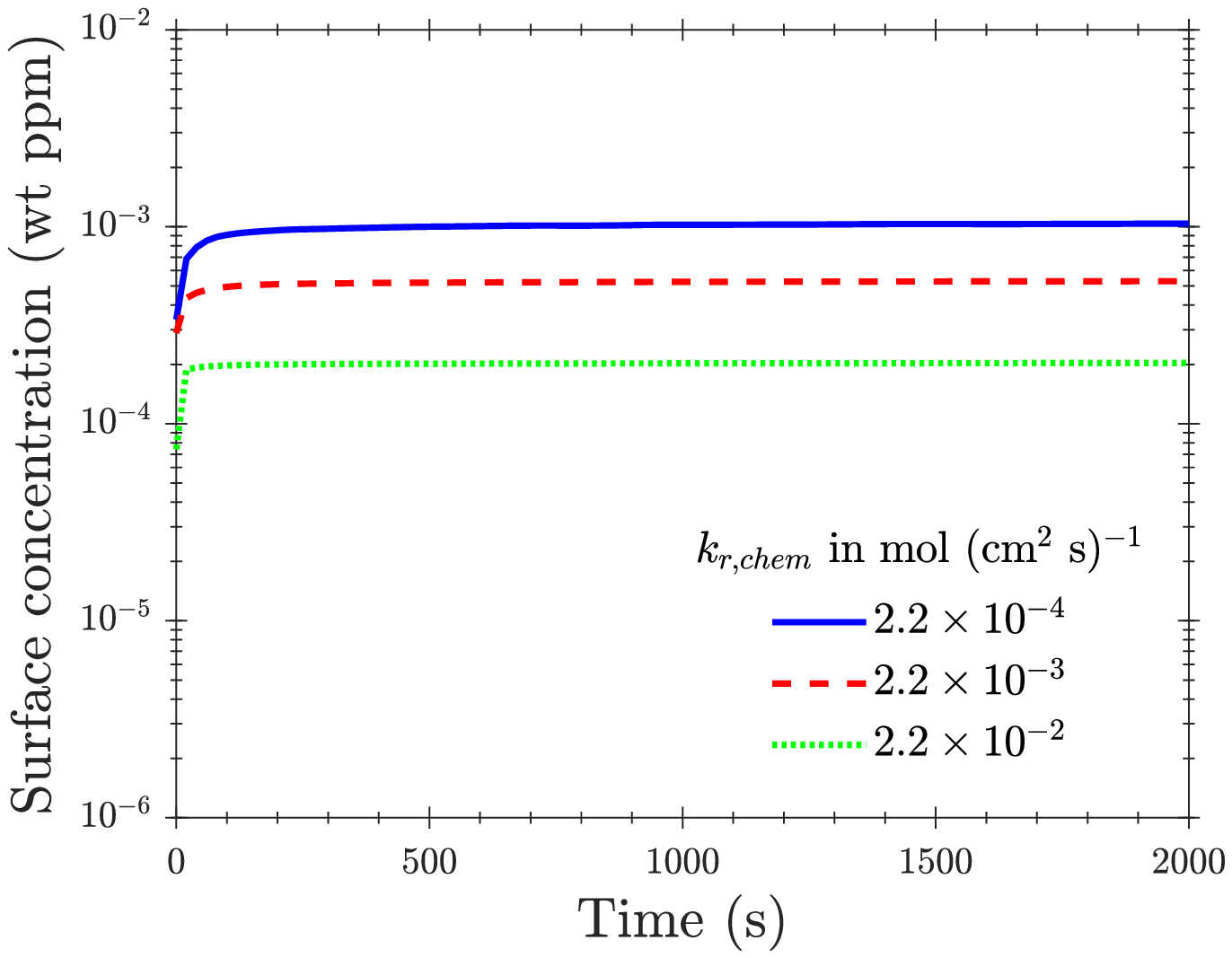}
                \caption{}
                \label{fig:Cs_krchem}
        \end{subfigure}
        \begin{subfigure}[b]{0.6\textwidth}
                \raggedleft
                \includegraphics[scale=0.55]{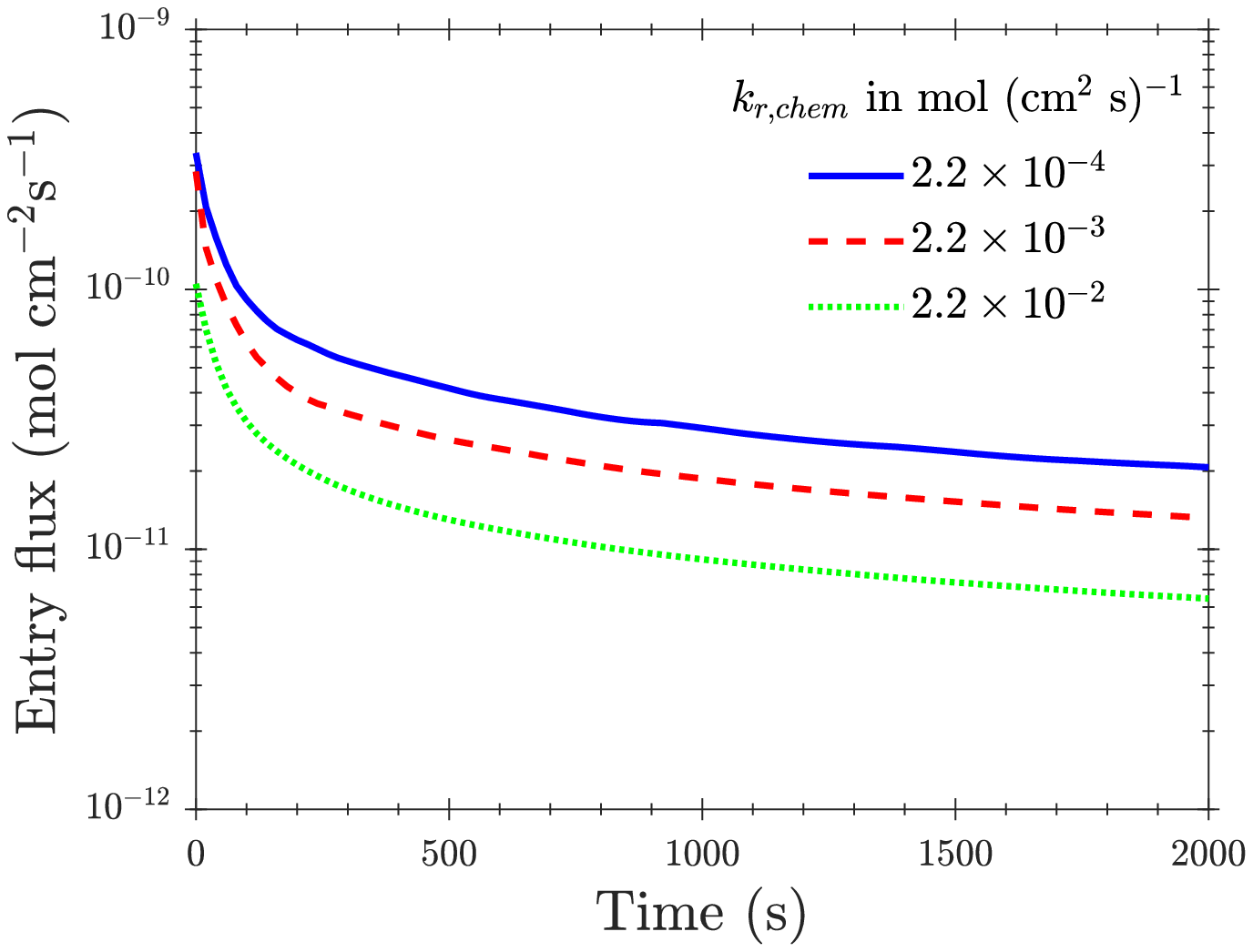}
                \caption{}
                \label{fig:Jin_krchem}
        \end{subfigure}}
        \caption{Influence of $k_{r,chem}$ on (a) sub-surface concentration and (b) hydrogen entry flux due to generalised boundary conditions}
        \label{fig:krchem}
\end{figure}

\begin{figure}[H]
\makebox[\linewidth][c]{%
        \begin{subfigure}[b]{0.6\textwidth}
                \centering
                \includegraphics[scale=0.55]{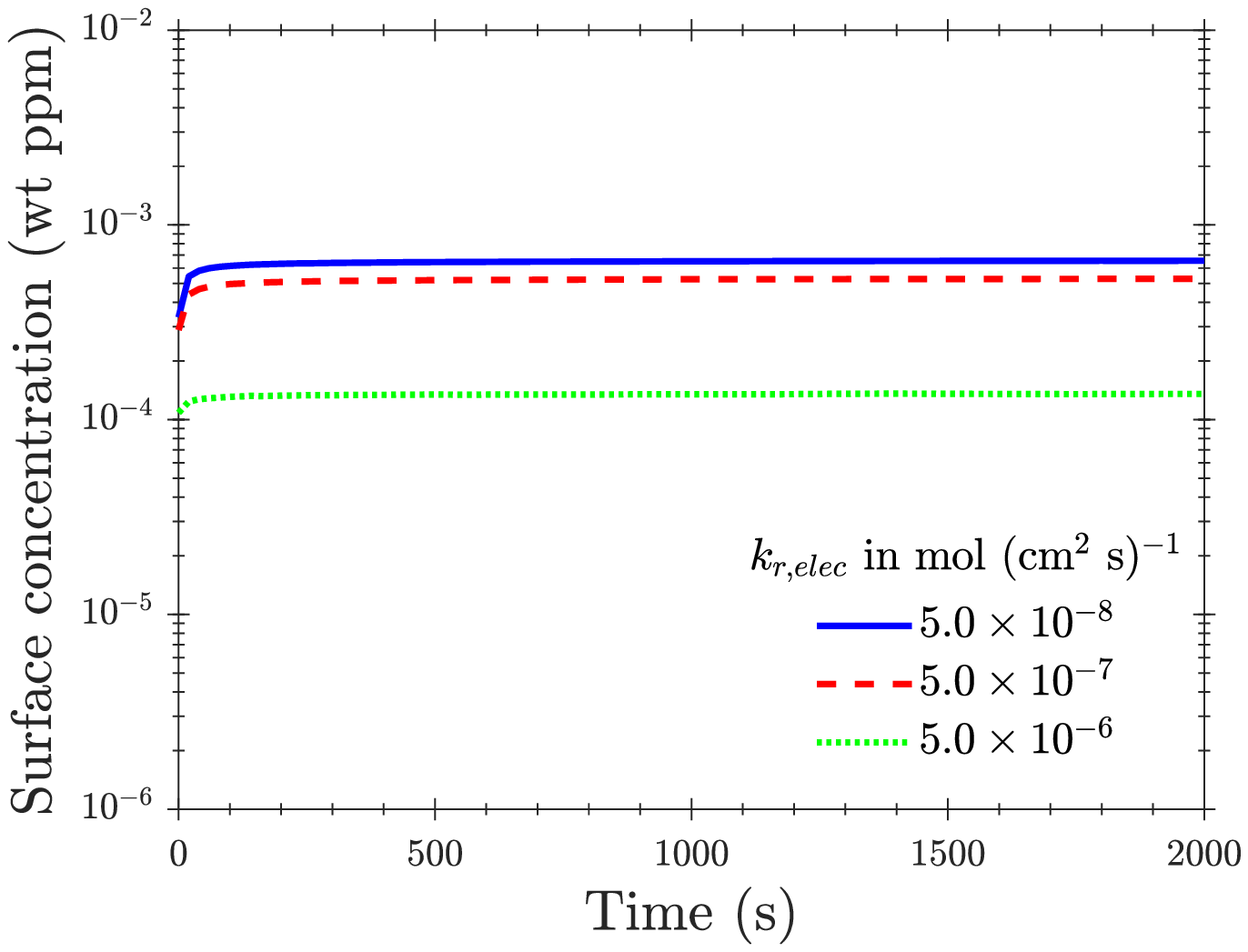}
                \caption{}
                \label{fig:Cs_krelec}
        \end{subfigure}
        \begin{subfigure}[b]{0.6\textwidth}
                \raggedleft
                \includegraphics[scale=0.55]{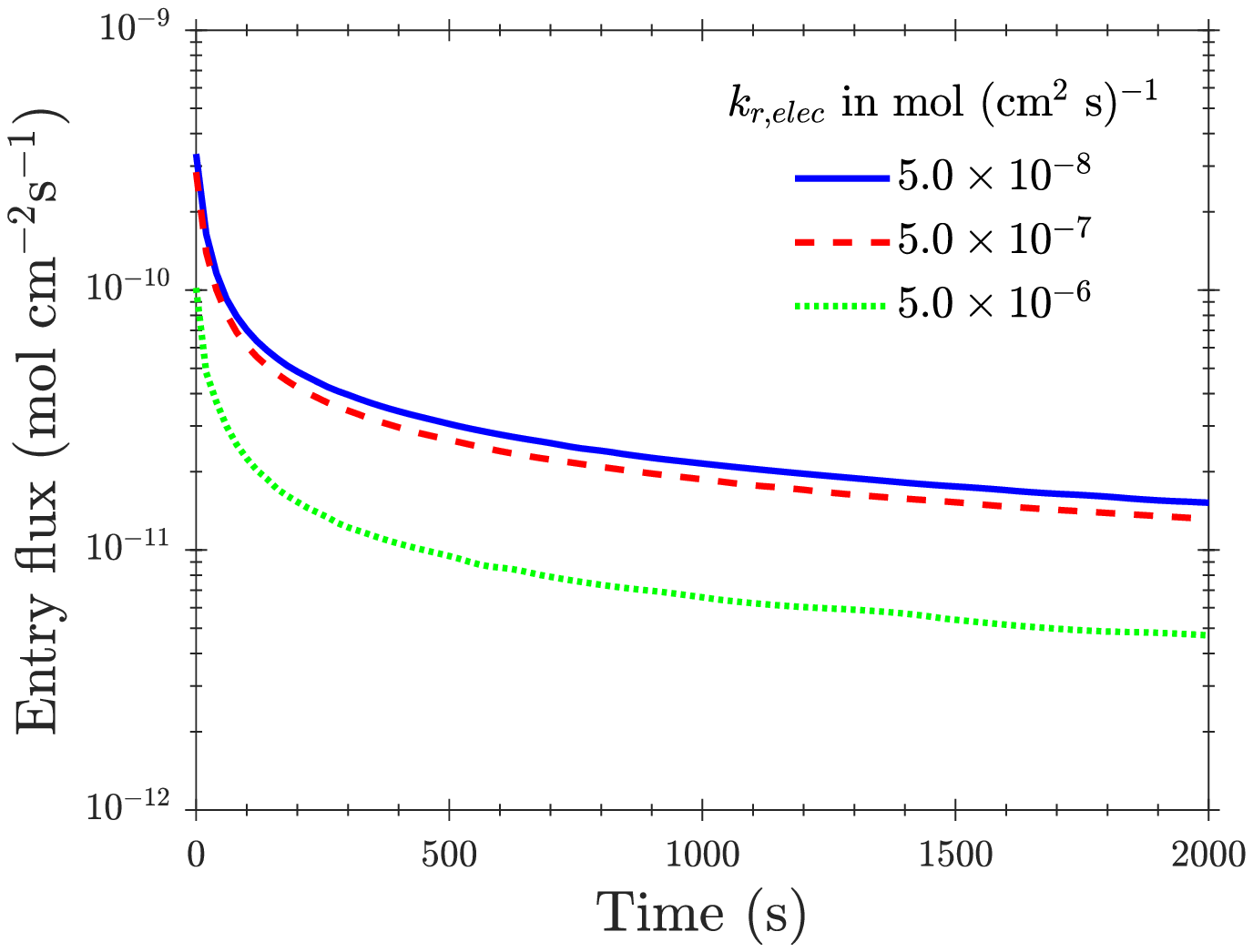}
                \caption{}
                \label{fig:Jin_krelec}
        \end{subfigure}}
        \caption{Influence of $k_{r,elec}$ on (a) sub-surface concentration and (b) hydrogen entry flux due to generalised boundary conditions}
        \label{fig:krelec}
\end{figure}

The influence of the charging constant $k_c$ is related to the input current density. Legend values shown in Fig. \ref{fig:kc} are equivalent to charging constants $k_c$ equal to $0.048$, $0.48$, and $4.8$ $A/m^{2}$. As expected, the higher current densities, the higher the entry fluxes and concentrations obtained. The inverse effect is found for the recombination constants (Figs. \ref{fig:krchem} and \ref{fig:krelec}), such that high values of $k_{r,chem}$ and $k_{r,elec}$ lead to a lower hydrogen uptake. It can be concluded that the competing absorption-desorption and charging/recombination processes will determine the amount of hydrogen that enters to the bulk material, and consequently they must be experimentally measured for different conditions. 

\bibliographystyle{elsarticle-num} 
\bibliography{library}


\end{document}